\documentclass{article}
\usepackage[verbose=true,letterpaper]{geometry}
\AtBeginDocument{
  \newgeometry{
    textheight=9in,
    textwidth=6.5in,
    top=1in,
    headheight=14pt,
    headsep=25pt,
    footskip=30pt
  }
}

\newcommand{\headeright}{A Preprint}

\usepackage{fancyhdr}
\fancyheadoffset{0pt}
\def\keywordname{{\bfseries \emph Keywords}}%
\def\keywords#1{\par\addvspace\medskipamount{\rightskip=0pt plus1cm
\def\and{\ifhmode\unskip\nobreak\fi\ $\cdot$}\noindent\keywordname\enspace\ignorespaces#1\par}}

\usepackage{authblk}
\usepackage{todonotes}
\usepackage[utf8]{inputenc} 
\usepackage[T1]{fontenc}    
\usepackage{hyperref}       
\usepackage{url}            
\usepackage{booktabs}       
\usepackage{amsfonts}       
\usepackage{nicefrac}       
\usepackage{microtype}      
\usepackage{lipsum}		    
\usepackage{graphicx}
\usepackage{times}
\usepackage{array}
\usepackage{amssymb, amsmath, amsthm}
\usepackage{graphicx}
\usepackage{lmodern,url}
\usepackage{makecell} 
\usepackage{cancel}
\usepackage{multirow}
\usepackage{microtype}
\usepackage{lineno}
\usepackage{xspace}
\usepackage{xcolor}
\usepackage{siunitx}
\usepackage{todonotes}
\usepackage{booktabs}
\usepackage{arydshln}
\usepackage{optidef}
\usepackage{mathdots}
\usepackage{subcaption}
\usepackage{csquotes}
\usepackage{caption}
\allowdisplaybreaks
\newcommand{\figref}[1]{Fig.~\ref{#1}}
\newcommand{\tabref}[1]{\tablename~\ref{#1}}

\newcommand{\dis}{\displaystyle}

\graphicspath{{Figures/}}

\newcommand{\Nhatobs}{\hat{N}^{\text{obs}}}
\newcommand{\xiap}{\xi^{\text{ap}}}

\newcommand{\xunderbrace}[2][\vphantom{\dfrac{A}{A}}]{\underbrace{#1#2}}
\newcommand{\EH}{E^{H}}

\newcommand{\ET}{E^{Q}}

\newcommand{\Ht}{I^{H}}
\newcommand{\Hta}{I^{H,a}}
\newcommand{\Hts}{I^{H,s}}
\newcommand{\Tt}{I^{Q}}
\newcommand{\Tta}{I^{Q,a}}
\newcommand{\Tts}{I^{Q,s}}
\newcommand{\ftestH}{N^{\rm test}}
\newcommand{\ftestHs}{N_{s}^{\rm test}}
\newcommand{\ftrace}{N^{\rm traced}}
\newcommand{\alphaT}{\left(\chisr\left(1\!-\!\xi\right) + \chir\xi\right)}
\newcommand{\SM}{\frac{S}{M}}
\newcommand{\Htlag}{\Ht_{t-\tau}}
\newcommand{\Hslag}{\Hts_{t-\tau}}
\newcommand{\Htsmax}{\Hts_{\rm max}}
\newcommand{\Htmax}{\Ht_{\rm max}}
\newcommand{\Ntest}{N^{\rm test}}
\newcommand{\Ntestlag}{\Ntest_{t-\tau}}
\newcommand{\Nmax}{\Ntest_{\text{max}}}  
\newcommand{\latRate}{\rho}
\newcommand{\Phit}{\Phi_t}
\newcommand{\DL}{D_{\rm L}}
\newcommand{\Dramp}{D_{\rm ramp}}
\newcommand{\chitau}{\chi_{\tau}}
\newcommand{\chisr}{\chi_{s,r}}
\newcommand{\chir}{\chi_{r}}

\newcommand{\RtT}{\left(\nu+\epsilon\right) R_0}
\newcommand{\RelContacts}{k_t}
\newcommand{\RtH}{\RelContacts R_0}

\newcommand{\Rtobs}{\hat{R}_t^\text{obs}}

\newcommand{\RelContactsCrit}{\RelContacts^{\rm crit}}

\newcommand{\RelContactsld}{k_{\rm LD}}

\newcommand{\RelContactsald}{k_{\rm nLD}}

\newcommand{\Nequil}{\hat{N}^\text{obs}_\infty}

\usepackage[compress]{scicite}
\title{Low case numbers enable long-term stable pandemic control without lockdowns}

\usepackage{authblk}

\author[1,2]{Sebastian Contreras}
\author[1]{Jonas Dehning}
\author[1]{Sebastian B.~Mohr}
\author[1]{Simon Bauer}
\author[1]{F.~Paul Spitzner}
\author[1,3*]{Viola Priesemann}

\affil[1]{Max Planck Institute for Dynamics and Self-Organization, Am Fa{\ss}berg 17, 37077 G\"ottingen, Germany.}
\affil[2]{Centre for Biotechnology and Bioengineering, Universidad de Chile, Beauchef 851, 8370456 Santiago, Chile.}
\affil[3]{Department of Physics, University of G\"ottingen, Friedrich-Hund-Platz 1, 37077 G\"ottingen, Germany.}
\affil[$\,$]{*Corresponding author. \tt{viola.priesemann@ds.mpg.de}}
\affil[$\,$]{All authors contributed equally.}

\date{}

\renewcommand{\headeright}{ }

\begin{document}
\maketitle
\begin{abstract} 



The traditional long-term solutions for epidemic control involve eradication or population immunity. Here, we analytically derive the existence of a third viable solution: a stable equilibrium at low case numbers, where test-trace-and-isolate policies partially compensate for local spreading events, and only moderate restrictions remain necessary. 
In this equilibrium, daily cases stabilize around ten new infections per million people or less.
However, stability is endangered if restrictions are relaxed or case numbers grow too high. The latter destabilization marks a tipping point beyond which the spread self-accelerates. 
We show that a lockdown can reestablish control and that recurring lockdowns are not necessary given sustained, moderate contact reduction. We illustrate how this strategy profits from vaccination and helps mitigate variants of concern.
This strategy reduces cumulative cases (and fatalities) 4x more than strategies that only avoid hospital collapse. 
In the long term, immunization, large-scale testing, and international coordination will further facilitate control.
\end{abstract} 

\section*{Introduction}
As SARS-CoV-2 is becoming endemic and knowledge about its spreading is accumulated, it becomes clear that neither global eradication nor population immunity will be achieved soon.
Eradication is hindered by the worldwide prevalence and by asymptomatic spreading. Reaching population immunity without an effective vaccine or medication would take several years and cost countless deaths, especially among the elderly \cite{randolph2020herd,Levin2020}. Moreover, evidence for long-term effects (``long COVID'') is surfacing~\cite{Alwan2020,Fraserm3001, greenhalgh_management_2020, topol_covid-19_2020}, advising against strategies aiming to progressively exposing people to the disease so that they acquire natural immunity. 
Hence, we need long-term, sustainable strategies to contain the spread of SARS-CoV-2. The common goal, especially in countries with an aging population, should be to minimize the number of infections and, thereby, allow reliable planning for individuals and the economy -- while not constraining individuals' number of contacts too much \cite{Priesemann2020panEur}. Intuitively, a regime with low case numbers would benefit not only public health and psychological well-being but would also profit the economy \cite{scherbina2020determining,xiong2020impact}.

However, control of SARS-CoV-2 is challenging.
Many infections originate from asymptomatic or pre-symptomatic cases \cite{li2020substantial} or indirectly through aerosols \cite{van2020aerosol}, rendering mitigation measures difficult.
Within test-trace-and-isolate (TTI) strategies, the contribution of purely symptom-driven testing is limited, but together with contact tracing, it can uncover asymptomatic chains of infections.
The rising availability of effective vaccines against SARS-CoV-2 promises to relieve the social burden caused by non-pharmaceutical interventions (NPIs). However, it is unclear how fast the restrictions can be lifted without risking another wave \cite{moore2021vaccination,bauer2021relaxing,contreras2021risking}, and how well vaccines will protect against more contagious or immune-response-escaping variants.
Additional challenges are the potential influx of SARS-CoV-2 infections (brought in by travelers or commuting workers from abroad), imperfect quarantine, limited compliance, and TTI and case-reporting delays.
Lastly, any country's capacity to perform TTI is limited, so that spreading dynamics change depending on the level of case numbers. Understanding these dynamics is crucial for informed policy decisions.

\section*{Results}
\subsection*{Analytical framework: overview}

We analytically show the existence of a stable regime at low case numbers, where control of SARS-CoV-2 is much easier to achieve and sustain. In addition, we investigate mitigation strategies and long-term control for COVID-19, where we build on our past work to understand the effectiveness of non-pharmaceutical interventions, particularly test-trace-and-isolate (TTI) strategies \cite{contreras2021challenges,Linden2020DAE,dehning2020inferring}. Importantly, the strategy we propose does not rely on the availability of a cure or vaccine, and it is applicable not only to further waves of COVID-19 but also to novel diseases with pandemic potential. Nonetheless, we also show how vaccination campaigns will further facilitate the success of the proposed strategy, assuming a vaccination rate as planned for countries in the European Union \cite{bauer2021relaxing}.

For quantitative assessments, we adapt an SEIR-type compartmental model~\cite{hethcote_mathematics_2000} to explicitly include a realistic TTI system that considers the challenges above. In our framework, individuals can be tested (and subsequently quarantined if tested positive) by three different mechanisms. First, symptomatic infections with COVID-19-specific symptoms would self-report or be diligently identified by surveillance and get a preferential test. Second, random asymptomatic screening would be homogeneously deployed in the general population disregarding symptoms so that every individual could be tested alike. Third, all the close contacts of those individuals recently tested positive by the two mechanisms mentioned above would also be tested. Understandably, limited resources pose a complex challenge for resource allocation, where efficient TTI would only be possible at low case numbers. We also built an interactive platform where enthusiastic readers can simulate scenarios different from those presented herein \url{http://covid19-metastability.ds.mpg.de/}. 

A central parameter for our analysis is the \textit{level of contagious contacts} $\RelContacts$ (relative to pre-COVID-19). More precisely, $\RelContacts$ refers to the fraction of infection-risk-bearing encounters compared to pre-COVID-19 contact levels. We can then interpret $\RelContacts$ in terms of the hidden reproduction number $R_t^{H}$, which accounts for the number of offspring infections generated by individuals unaware of being infectious, i.e., hidden infections in a naive and fully susceptible population \cite{contreras2021challenges}. Thus, in terms of $R_t^{H}$, $\RelContacts$ can be understood as the ratio between the offspring infections a hidden individual would generate in the presence and in the absence of NPIs, in other words, $R_t^{H}=\RelContacts R_0$.
Apart from direct contact reduction, contributions that allow increasing $\RelContacts$ without compromising the stability of the system also come from improved hygiene, mandatory face-mask policies, frequent ventilation of closed spaces, and avoiding indoor gatherings, among other precautionary measures.
As the latter measures are relatively fixed, direct contact reduction remains the central free variable, which is also the one tuned during lockdowns. All other parameters (and their references) are listed in \tabref{tab:Parametros}. 

\subsection*{Equilibrium at low case numbers}

\begin{figure}[!h]
    \centering
    \includegraphics[width =15cm]{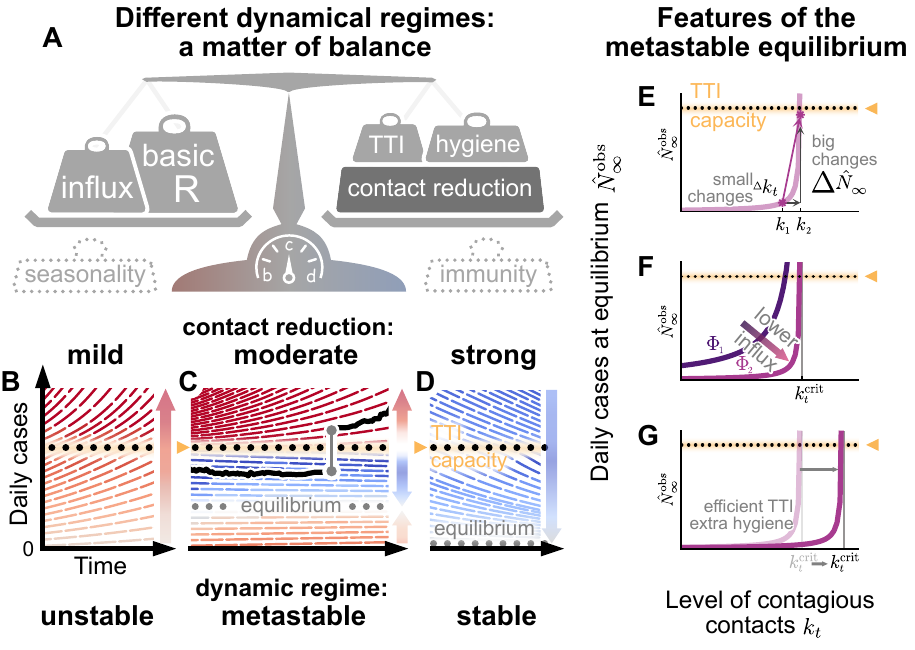}
    \caption{\textbf{Spreading dynamics depend on the balance between destabilizing and stabilizing contributions and on the level of case numbers.}
        \textbf{A:} Among the factors that destabilize the spread, we find the basic reproduction number $R_0$, and the external influx of infections (and possibly seasonality).
        On the other hand, increased hygiene, test-trace-isolate (TTI) strategies, contact reduction, and immunity contribute to stability.
        We specifically investigated how reductions in the contact level $\RelContacts$ and limited TTI capacity determine the stabilization of case numbers.
        \textbf{B:} At mild contact reduction ($\RelContacts=\SI{80}{\percent}$ compared to pre-COVID-19 times), TTI is not sufficient; case numbers would grow even when TTI capacity is available.
        \textbf{C:} At moderate contact reduction ($\RelContacts=\SI{60}{\percent}$), a metastable equilibrium emerges (gray dots) to which case numbers converge, if they were below the TTI capacity.
        However, destabilizing events (e.g., a sudden influx of infections) can push a previously stable system above the TTI capacity and lead to an uncontrolled spread (black line as an example).
        \textbf{D:} Assuming strong contact reduction ($\RelContacts=\SI{40}{\percent}$), case numbers decrease even if the TTI capacity is exceeded.
        \textbf{E:} Near to the critical level of contacts $\RelContactsCrit$, small changes in the contact level will lead to a considerable increase in observed cases in equilibrium $\Nequil$. 
        \textbf{F:} Reducing the influx of infections $\Phit$ (by closing borders or deploying extensive testing at arrival) reduces the number of infections.
        \textbf{G:} Increasing the efficiency of manual contact tracing and additional measures as increased hygiene and compulsory use of face masks will increase the maximum allowed level of contacts $\RelContactsCrit$. 
        }
    \label{fig:abstract}
\end{figure}

We find a regime where the spread reaches an equilibrium at low daily case numbers between the scenarios of eradication of the disease or uncontrolled spreading. The main control parameter that determines whether the system can reach an equilibrium is the level of contagious contacts $\RelContacts$. 

If the reduction in the contact level $\RelContacts$ is mild, case numbers grow exponentially, as measures could not counterbalance the basic reproduction number ($R_0 \approx 3.3$ for SARS-CoV-2~\cite{ZHAO2020214,alimohamadi2020151}) (\figref{fig:abstract}B).
In contrast, if the reduction in $\RelContacts$ is strong and (together with hygiene and TTI) outweighs the drive by the basic reproduction number, case numbers decrease to a low equilibrium value (\figref{fig:abstract}D). 

Importantly, if the reduction in $\RelContacts$ is moderate (and just-about balances the drive by the basic reproduction number), we find a metastable regime:
The spread is stabilized if and only if the overall case numbers are sufficiently low to enable fast and efficient TTI (\figref{fig:abstract}C).
However, this control is lost if the limited TTI capacity is overwhelmed. Beyond that tipping point, the number of cases starts to grow exponentially as increasingly more infectious individuals remain undetected \cite{contreras2021challenges}. 

The capacity of TTI determines the minimal required contact reduction for controlling case numbers around an equilibrium.
If case numbers are sufficiently below the TTI capacity limit, the maximum allowed level of contacts to enable the (meta-)stable regime in our default scenario (cf.~\tabref{tab:Parametros}, with $R_0=3.3$) is $\RelContactsCrit = \SI{61}{\%}$ (95\% confidence interval (CI):[\SI{47}{}, \SI{76}{}]). When the level of contacts $\RelContacts$ is below the threshold, case numbers asymptotically approach to an equilibrium that shows the following features: i) when the level of contacts $\RelContacts$ is close to its critical value $\RelContactsCrit$, small changes in $\RelContacts$ generate large modifications of the equilibrium level~(\figref{fig:abstract}E). ii) Larger influxes lead to larger equilibrium values, however not modifying the maximum allowed level of contacts~(\figref{fig:abstract}F). iii) Behavioral changes and policies leading to a reduction in the transmission probability will also lower the equilibrium value because the maximum allowed level of contacts $\RelContactsCrit$ would be larger~(\figref{fig:abstract}G).

However, if case numbers exceed the TTI capacity limit, a considerably stronger reduction in contacts is required to reach the stable regime, so that $\RelContactsCrit = \SI{42}{\%}$ (95\% CI: [\SI{38}{}, \SI{47}{}])  (\figref{fig:tipping_point}~B, Fig.~S2B and Table~S1). 

\subsection*{Equilibrium depends on influx and contact reduction}\label{sec:EqValues}

\begin{figure}[!h]
    \centering
    \includegraphics[width = 12cm]{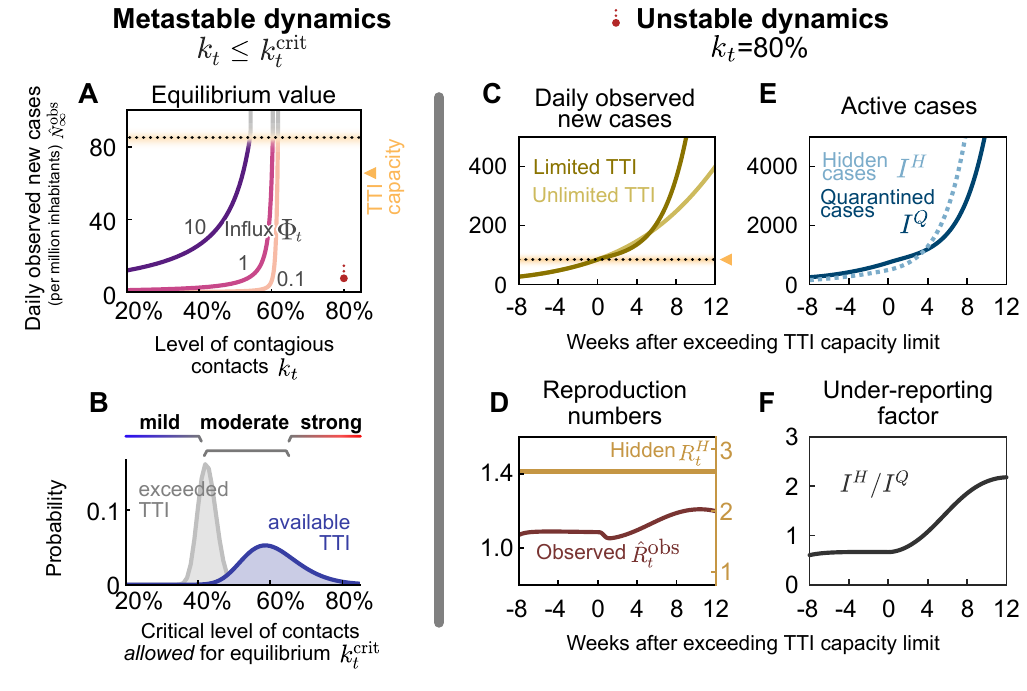}
    \caption{%
        \textbf{(A, B) In the stable and metastable regimes, daily new cases approach an equilibrium value $\Nequil$ that depends on contact level $\RelContacts$ and the external influx of new cases $\Phit$.}
        \textbf{A:}
        The equilibrium value $\Nequil$ increases with increased contact level $\RelContacts$ or higher influx $\Phit$.
        No equilibrium is reached if either $\RelContacts$ or $\Phit$ are above the respective (critical) threshold values.
        \textbf{B:} The critical value $\RelContactsCrit$ represents the maximal contact level that is allowed to reach equilibrium and stabilize case numbers.
        If case numbers are below the TTI capacity limit, higher values of $\RelContactsCrit$ are permitted for stabilization (blue) than if cases exceed TTI (gray). Confidence intervals originate from error propagation of the uncertainty of the underlying model parameters.
        \textbf{(C--F)
        In the unstable dynamic regime ($\RelContacts=80\%$), a tipping point is visible when exceeding TTI capacity.}
         We observe a self-accelerating increase of case numbers after crossing the TTI limit (\textbf{C}) and a subsequent increase of the reproduction number (\textbf{D}).
         Furthermore, the absolute number (\textbf{E}) and the fraction (\textbf{F}) of cases that remain unnoticed increase over time.} 
    \label{fig:tipping_point}
\end{figure}

If an equilibrium is reached, the precise value of daily new cases $\Nequil$ at which the system stabilizes depends on both the contact level ($\RelContacts$) and the external influx of new cases ($\Phit$) (\figref{fig:tipping_point}A). In general, for realistic low values of influx $\Phit$, the equilibrium level $\Nequil$ is low. 
However, $\Nequil$ increases steeply (diverges) when the contact level $\RelContacts$ approaches the tipping point to unstable dynamics (\figref{fig:tipping_point}A,B). Such a divergence near a critical point $\RelContactsCrit$  is a general feature of continuous transitions between stable and unstable dynamics~\cite{lasalle1976stability,Wilting2018}. As a rule of thumb, in an analytical mean-field approximation~\cite{Wilting2018}, $\Nequil$ would be proportional to $\Phit$ and diverge when $\RelContacts$ approaches its critical value $\RelContactsCrit$ from below: 
\begin{equation}
    \Nequil \propto \frac{\Phit}{\RelContactsCrit-\RelContacts}.
\end{equation}

Robust control of the pandemic requires maintaining a sufficient safety margin from the tipping point (and the subsequent transition to instability) for two reasons. 
First, small fluctuations in $\RelContacts$, $\Phit$ (or other model variables) could easily destabilize the system.
Second,
near the critical value $\RelContactsCrit$,
reductions in $\RelContacts$ are especially effective: already small further reductions below $\RelContactsCrit$ lead to significantly lower stable case numbers (\figref{fig:tipping_point}A).
Already with moderate reductions in $\RelContacts$ ($\SI{50}{\percent} < \RelContacts < \SI{60}{\percent}$), the spread can be stabilized to a regime of case numbers clearly below 10 per million (Fig.~S1B, lower right region). 

\subsection*{Limited TTI and self-acceleration}\label{sec:tippingpoint}

If mitigation measures are insufficient, case numbers rise and eventually surpass the TTI capacity limit. 
Beyond it, health authorities cannot efficiently trace contacts and uncover infection chains. Thus the control of the spread becomes more difficult. We start our scenario with a slight increase in the case numbers over a few months, as seen in many European countries throughout summer 2020 (Fig.~S4 and Fig.~S5).
A tipping point is then visible in the following observables (\figref{fig:tipping_point}~C--F):

First, when case numbers surpass the TTI capacity, the increase in daily new observed cases $\Nhatobs$ becomes steeper, growing even faster than the previous exponential growth (\figref{fig:tipping_point}~C, full versus faint line).
The spread self-accelerates because increasingly more contacts are missed, which, in turn, infect more people. Importantly, in this scenario, the accelerated spread arises solely because of exceeding the TTI limit --- without any underlying behavior change among the population.

Second,
after case numbers surpass the TTI limit,
the observed reproduction number $\Rtobs$, which had been only slightly above the critical value of unity, increases significantly by about \SI{20}{\percent} (\figref{fig:tipping_point}~D). This reflects a gradual loss of control over the spread and explains the faster-than-exponential growth of case numbers.
The initial dip in $\Rtobs$ is a side-effect of the limited testing: As increasingly many cases are missed, the observed reproduction number reduces transiently. 

Third, compared to the infectious individuals who are quarantined $\Tt$,
the number of infectious individuals who are hidden $\Ht$ (i.e. those who are not isolated or in quarantine) increases disproportionately (\figref{fig:tipping_point}~E) which is measured by the \textquote{under-reporting factor} ($\Ht / \Tt$) (\figref{fig:tipping_point}~F).
The hidden infectious individuals are the silent drivers of the spread as they, unaware of being infectious, inadvertently transmit the virus. This implies a considerable risk, especially for vulnerable people.
The TTI system can compensate for the hidden spread at low case numbers because it uncovers hidden cases through contact tracing.
However, at high case numbers, the TTI becomes inefficient:
If the TTI measures are \textquote{slower than the viral spread},
many contacts cannot be quarantined before they become spreaders.

\subsection*{Re-establishing control with lockdowns}\label{sec:LockdownParameters}

\begin{figure}[!h]
    \centering
    \includegraphics[width=16cm]{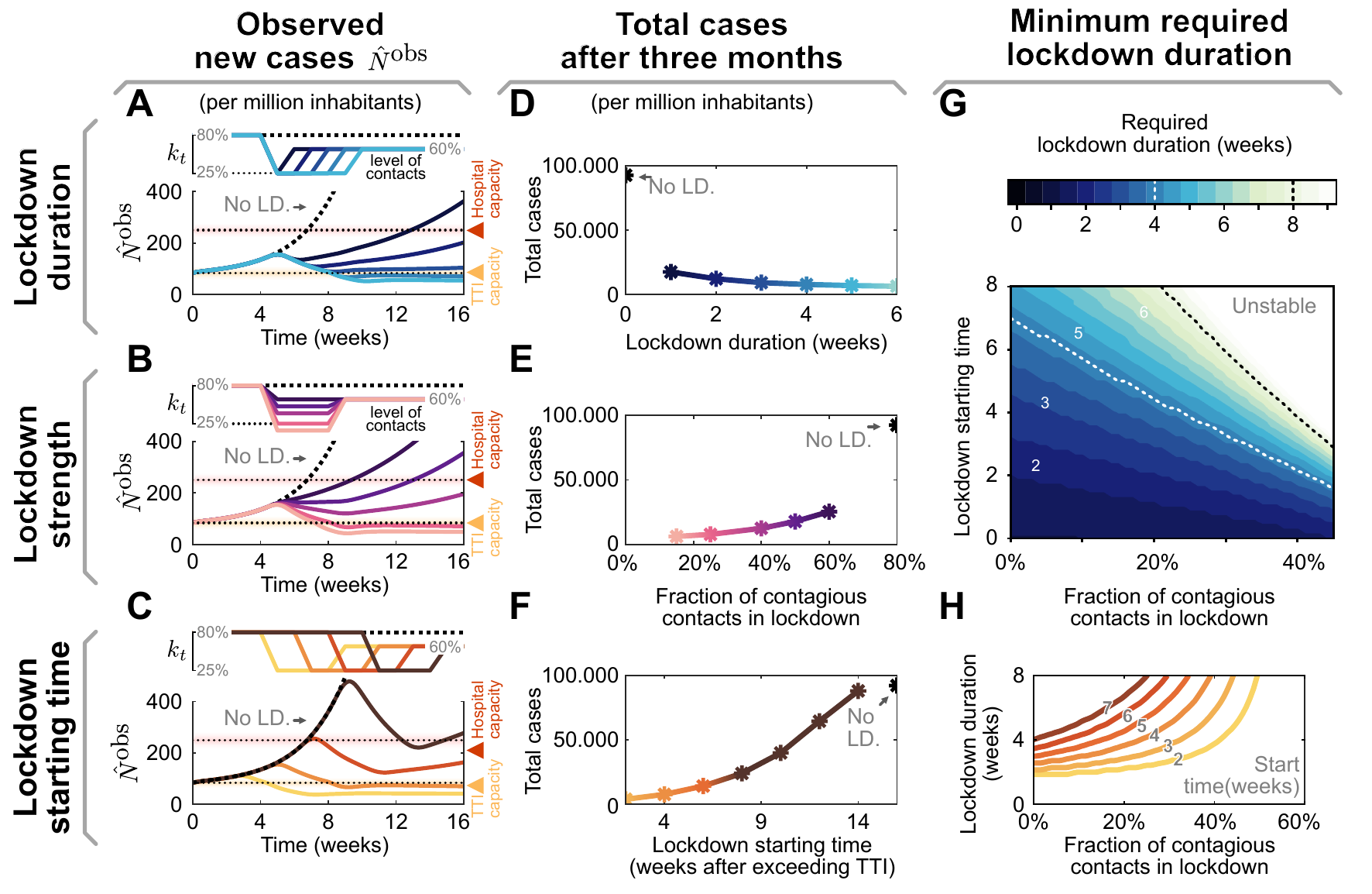}
    \caption{\textbf{The effectiveness of a lockdown depends on three primary parameters: its duration, stringency (strength), and starting time.}
    (\textbf{A--C}) 
    Observed daily new cases for a lockdown (abbreviated as LD), enacted after exceeding the TTI capacity. Reference parameters are a lockdown duration of 4 weeks, reducing contact level to $\RelContacts = 25\,\%$ and a start time at 4 weeks after exceeding TTI capacity. We vary lockdown duration from 1 to 5 weeks (\textbf{A}), lockdown strength (\textbf{B}) and lockdown starting time (\textbf{C}) to investigate whether stable case numbers can be reached.
    \textbf{D--F:}
    Total cases after three months, if the lockdown is parameterized as described in panels A--C, respectively.
    \textbf{G,~H:}~The minimal required duration of lockdown to reach equilibrium depends both on strength and start time. \textbf{G:} Heavy contact reduction (leading to lower values of $\RelContacts$) and timely lockdown enacting can create effective short lockdowns ($\leq$ 2 weeks, lower left, dark region). Whereas with mild contact reduction and very late start times, lockdowns become ineffective even when they last indefinitely (upper right, bright area).
    \textbf{H:} Horizontal slices through the color-map (\textbf{G}). Here, colors match panels (C,~F) and correspond to the lockdown start time.}
    \label{fig:parameters}
\end{figure} 

Once the number of new infections has overwhelmed the TTI system, re-establishing control can be challenging. 
A recent suggestion is the application of a circuit breaker \cite{keeling_precautionary_2020,mahase_covid-19_2020,kmietowicz_covid-19_2020}, a short lockdown to significantly lower the number of daily new infections.
Already during the so-called "first wave" of COVID-19 in Europe (i.e., the time-frame between March and June, 2020), lockdowns have proven capable of lowering case numbers by a factor of 2 or more every week \cite{Brauner2020EffectivenessEurope,dehning2020inferring} (corresponding to an observed reproduction number of $\Rtobs \approx 0.7$).  With the knowledge we now have acquired about the spreading of SARS-CoV-2, more targeted restrictions may yield a similarly strong effect. 

Inspired by the lockdowns installed in many countries\cite{li_temporal_2020}, we assume a default lockdown of four weeks,  starting four weeks after case numbers exceed the TTI capacity limit, and a strong reduction of contagious contacts during a lockdown, leading to $\RelContacts=\SI{25}{\percent}$ (which corresponds to an $\Rtobs \approx 0.85$, see~Table~S2). We further assume that during lockdown, the external influx of infections $\Phit$ is reduced by a factor of ten and that after the lockdown, a moderate contact reduction (allowing $\RelContactsald=\SI{60}{\percent}$) is maintained. By varying the parameters of this default lockdown, we show in the following that the lockdown strength, duration, and starting time determine whether the lockdown succeeds or fails to reach equilibrium. 

In our scenario, a lockdown duration of four weeks is sufficient to reach the stable regime (\figref{fig:parameters}~A).
However, if lifted too early (before completing four weeks), case numbers will rise again shortly afterward. 
The shorter an insufficient lockdown, the faster case numbers will rise again.
Also, it is advantageous to remain in lockdown for a short time even after case numbers have fallen below the TTI limit --- to establish a safety margin, as shown above. 
Overall, the major challenge is not to ease the lockdown too early; otherwise, the earlier success is soon squandered.

During a lockdown, it is necessary to severely reduce contagious contacts to decrease case numbers below the TTI capacity limit (\figref{fig:parameters}~B). 
In our scenario, the contact level has to be reduced to at least $\RelContactsld=\SI{25}{\percent}$ to bring the system back to equilibrium.
Otherwise, a lockdown that is slightly weaker would fail to reverse the increasing trend in cases.
Furthermore, increasing the lockdown strength decreases both the required lockdown duration (\figref{fig:parameters}~G, H) and the total number of cases accumulated over three months (\figref{fig:parameters}~E). 
This shows that stricter lockdowns imply shorter-lasting social and economic restrictions.


The earlier a lockdown begins after exceeding the TTI capacity limit, the faster control can be re-established, and constraints can be loosened again (\figref{fig:parameters}~C). 
If started right after crossing the threshold, in principle, only a few days of lockdown are necessary to bring back case numbers below the TTI capacity limit.
On the other hand, if the lockdown is started weeks later,
its duration needs to increase (\figref{fig:parameters}~C,D) and the total number of cases will be significantly larger (\figref{fig:parameters}~F). The parameter regime between these two options is relatively narrow; it is not likely that equilibrium can eventually be reached as a lockdown exceeds many weeks (cf.~\figref{fig:parameters}~H).
For practical policies, this means that if a lockdown does not start to show apparent effects after 2 or 3 weeks, then the strategy should be revised (this assessment time is necessary due to the delay of 1-2 weeks between contagion and case report).

\subsection*{Maintaining control without lockdowns}\label{sec:longterm}

We show that repeated lockdowns are not required to maintain control over the COVID-19 spread if moderate contact reduction is maintained once case numbers are below the TTI capacity limit (an initial lockdown might still be necessary to establish control). 
A natural goal would be to keep case numbers below the hospital capacity.
However, our model suggests that lowering them below TTI capacity requires fewer contact restrictions (in the long-term), involves a shorter total lockdown duration, and costs fewer lives. 
In the following, we compare the long-term perspective of these two goals and their dependence on the necessary contact reduction.

\begin{figure}[tbh]
    \centering
    \includegraphics[width=16cm]{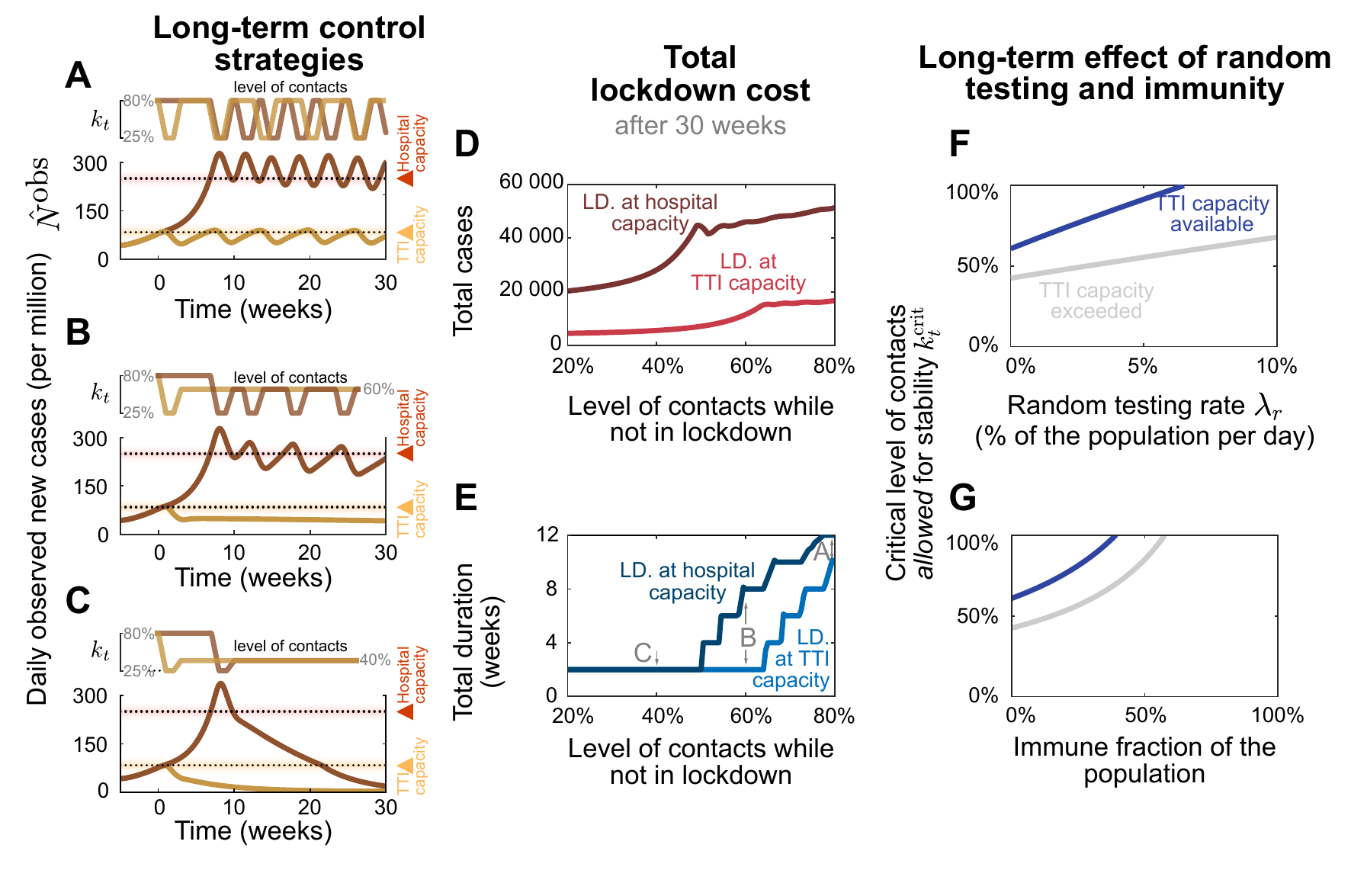}
    \caption{
    \textbf{From a long-term perspective, recurrent lockdowns are not required if the subsequent contact levels $\RelContacts$ are sufficiently low to reach equilibrium.} 
    A two-week lockdown of default strength (reducing contact level to $\RelContacts=\SI{25}{\%}$ relative to pre-COVID levels) is either enacted when the test-trace-and-isolate (TTI, yellow lines) or the hospital (red lines) capacity limits are crossed. 
    \textbf{A:}
    Given only a mild contact reduction while not in lockdown (to $\RelContacts=\SI{80}{\%}$), the imposed lockdowns need to reoccur with a high frequency. Moreover, the frequency at which they reoccur is higher at high case numbers (aiming at hospital capacity) than when aiming at the TTI capacity limit because of the self-accelerating effect described previously.
    \textbf{B:}
    If the contact reduction remains moderate (to $\RelContacts=\SI{60}{\%}$), lockdowns only need to reoccur if they fail to bring case numbers below TTI capacity. 
    \textbf{C:}
    Only lasting strong contact reduction (to $\RelContacts=\SI{40}{\%}$) is sufficient to continue to bring down case numbers after the lockdown.
    \textbf{D,~E:}
    The total cost (in terms of cumulative cases and total lockdown duration) depends on the level of contact reduction and the lockdown policy: The cost is low if lockdowns are initiated at the TTI capacity limit and if the contact level $\RelContacts$ after the lockdown is low.
    \textbf{F,~G:} Large-scale random testing (\textbf{F}) or immunization of the population (either post-infection or through vaccination \textbf{G}) increase the maximal contact level allowed for stability $\RelContactsCrit$.
    }
        \label{fig:circuit_breaker}
\end{figure} 

In our scenario (\figref{fig:circuit_breaker}), we start from the unstable regime, where the initial contact level ($\RelContacts=\SI{80}{\%}$ of the pre-COVID level) is not sufficient to control the spread. We start a two-week lockdown when crossing either the TTI or the hospital capacity.
During the lockdown, contacts are reduced to $\RelContacts = \SI{25}{\percent}$.
After the first and all subsequent lockdowns, contacts $\RelContacts$ are reduced to \SI{80}{\%},  \SI{60}{\%} or \SI{40}{\%} relative to pre-COVID-19 levels, thus representing a mild, moderate, or strong reduction. When assuming mild contact reduction after lockdowns, case numbers rise after lifting the lockdown, independent of the chosen threshold (TTI or hospital capacity, \figref{fig:circuit_breaker}~A). Thus, repeated lockdowns are necessary.


However, maintaining a moderate contact reduction while not in lockdown (i.e., a contact level of $\RelContacts =\SI{60}{\percent}$)
is sufficient to stay within the metastable regime --- 
if lockdowns are enacted such that case numbers remain below the TTI capacity (\figref{fig:circuit_breaker}~B, yellow line).
This is a promising perspective for a long-term control strategy that avoids recurrent lockdowns. 
Otherwise, if case numbers are above TTI capacity limit but below hospital capacity, control of the pandemic requires repeated lockdowns  (\figref{fig:circuit_breaker}~B, red line). Alternatively, lasting strong contact reductions even after the lockdown can be sufficient to drive down case numbers (\figref{fig:circuit_breaker}~C).

The advantage of the strategy to stay below the TTI capacity limit becomes very clear when considering the total cost of the required lockdowns:
Independent of the degree of contact reduction, (i) the total number of cases (and consequently deaths and \textquote{long-COVID} risk) is lower (\figref{fig:circuit_breaker}~D), (ii) the total duration spent in lockdown is shorter (\figref{fig:circuit_breaker}~E), and (iii) the frequency at which lockdowns have to reoccur -- should the after-lockdown contact reduction not be enough to grant metastability -- is lower (\figref{fig:circuit_breaker}~A).
As case numbers and lockdown duration indicate economic costs, a strategy that respects the TTI limit offers a low economic toll,  enables mid-term planning  and provides trust to people and society.

The scalability of random testing (screening)~\cite{holt_slovakia_2020, mina_rethinking_2020, larremore_test_2020} and immunization programs play a critical role in long-term strategies;  both will increase the maximum level of contacts allowed ($\RelContactsCrit$) to maintain control (\figref{fig:circuit_breaker}~F,~G).   
Early effects of immunity can be seen in our scenario of the system stabilized at hospital capacity: The need for lockdowns becomes less frequent over time (\figref{fig:circuit_breaker}~A,~B red lines).
However, acquiring natural immunity comes at the cost of a prolonged high level of case numbers, subsequent ``long COVID'' cases~\cite{Fraserm3001,greenhalgh_management_2020, topol_covid-19_2020} and deceased people.
Whereas the duration of the immunity is yet unknown \cite{kirkcaldy2020covid}, this phenomenon still shows that immunity effects play an increasing role as model predictions extend further into the future. 

\subsection*{Vaccination greatly facilitates containment in the effective TTI regime}

In the following, we show how growing immunity granted by vaccination programs further facilitates both reaching low case numbers below TTI capacity and the stable control thereof. For quantitative assessments, we studied the effect of COVID-19 vaccination programs and how they affect the two control strategies discussed in the previous section.

Investigating explicit vaccination scenarios, we assume that 80\% of people getting offered vaccination accept this offer. Since, as of now, none of the available vaccines has been approved for people younger than 16 years, this corresponds to a vaccine uptake of roughly 70\% of the overall population (for European demographics). We model the delivery and administration of all doses to be completed within 32 weeks \cite{bauer2021relaxing}, which is comparable to the increasing vaccination rate in the European Union. We investigate two scenarios: that the average vaccine efficacy against transmission is 80\%, in line with reported values for current vaccines against the wild type and the widely dominant B.1.1.7 variant \cite{thompson2021interim,dagan2021BNT,voysey2021safety}, or 40\%, a possible scenario for partially immune-escaping variants.

\begin{figure}[tbh]
    \centering
    \includegraphics[width=16cm]{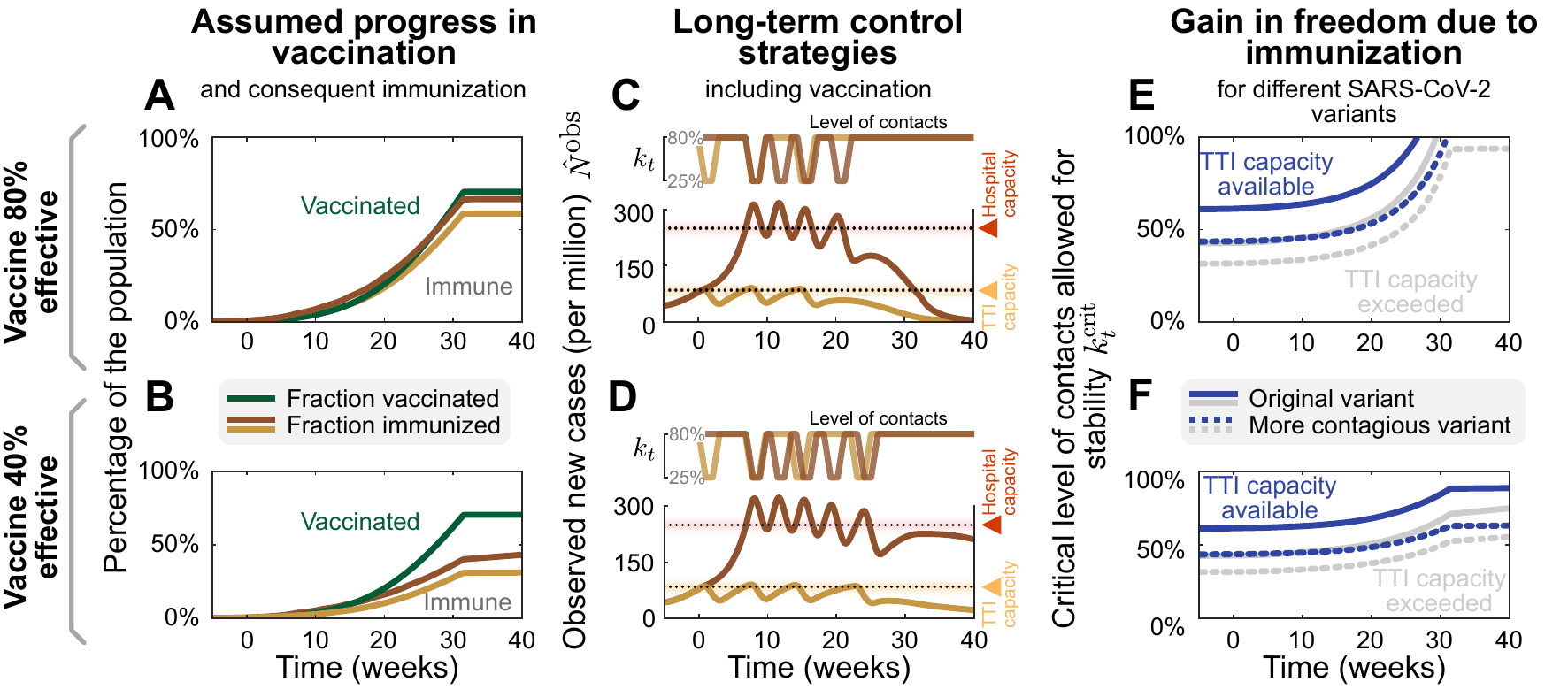}
    \caption{
    \textbf{Growing immunity due to mass vaccination will lead to declining case numbers and enable increasing contact levels} For 80\% (upper row) and 40\% (lower row) vaccine efficacy against transmission, we explore the effect of mass vaccination. \textbf{A,~B:} Vaccine rollout is assumed to accelerate over 32 weeks, ending when  70\% of the population have been vaccinated. \textbf{C,~D:} As in~\figref{fig:circuit_breaker}~A, a two-week lockdown of default strength (allowing only a $\RelContacts=\SI{25}{\%}$ of pre-COVID contacts) is either enacted when the test-trace-and-isolate (TTI, yellow lines) or the hospital (red lines) capacity limits are reached. While not in lockdown, people maintain a level of contacts of $\RelContacts=\SI{80}{\%}$. Without a major change in these policies, repeated lockdowns will eventually cease to be necessary due to the growing immunity, and case numbers will decline in all scenarios. However, this decline would occur weeks earlier if case numbers are below TTI capacity and vaccine efficacy is high. \textbf{E,~F:}  Growing immunity among the population will considerably increase the maximal --critical-- level of contacts allowed for stabilization of case numbers for both strategies, but more substantially if not exceeding TTI capacity limit. More contagious variants like B.1.1.7 (with an assumed base reproduction number of 4.3) will require a lower level of contacts (dotted lines). 
    }
        \label{fig:vaccination}
\end{figure}

Comparing the two different control strategies introduced in the preceding section, i.e., either aiming to keep case numbers below the TTI or below the hospital capacity limit, the progressing vaccination \figref{fig:vaccination}~A, B) will eventually lead to declining case numbers \figref{fig:vaccination}~C, D). However, this point will be reached several weeks earlier if case numbers are kept below TTI capacity. Until then, repeated lockdowns will remain necessary if contact reductions outside of lockdown are insufficient (see discussion in the preceding section). A decreasing number of infections will even be reached months earlier if TTI capacity is available than if it is overwhelmed. 

Low case numbers are greatly beneficial even in light of progressing vaccination programs.
In the high-case-number regime, only in the most optimistic scenario the spread of SARS-CoV-2  can be controlled without contact reductions after the vaccination program is finished (\figref{fig:vaccination}~E, the full gray line reaches 100\%). In all other scenarios, because of the dominance of more contagious (dotted lines) or immune escape variants lowering vaccine efficacy (lower row), an efficiently-functioning TTI program is necessary to allow for a high level of contacts (\figref{fig:vaccination}~E, F, blue lines). Therefore reaching low case numbers is complementary to the vaccination efforts and necessary to maximize the population's freedom.

\section*{Discussion}

We demonstrated that between the two extremes of eradication and uncontrolled spread, a metastable regime of SARS-CoV-2 spreading exists. In such a regime, every person only has to reduce their contacts moderately. Simultaneously, case numbers can still be maintained robustly at low levels because the test-trace-and-isolate (TTI) system can operate efficiently. If this regime is within reach, keeping case numbers below TTI capacity is a suitable strategy for the long-term control of COVID-19 (or other infectious diseases) that features low fatalities and a small societal burden. In addition, it maximizes the effect of large-scale pharmaceutical interventions (as vaccination programs).


\begin{figure}[!h]
    \centering
    \includegraphics[width=15cm]{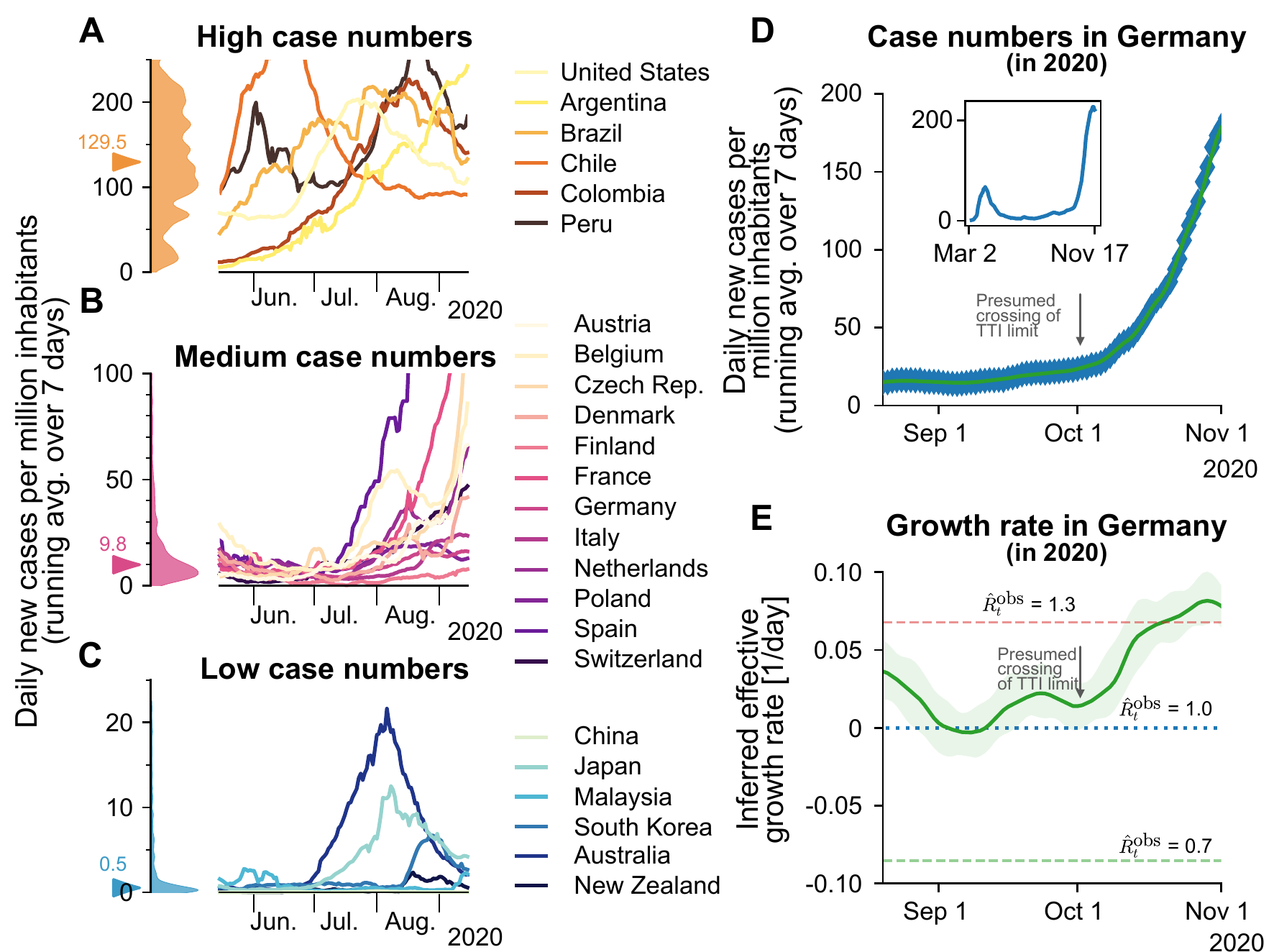}
    \caption{%
        \textbf{Strategies of countries to fight SARS-CoV-2 differ widely and are reflected in case numbers.}
        \textbf{A:}
        Strategies that involve little non-pharmaceutical interventions rely on the population to hinder the spread in a self-regulated manner and are often accompanied by high case numbers. 
        \textbf{B:}
        Strategies that aim to keep case numbers low through extensive TTI, combined with temporary lockdowns, can lead to medium case numbers.
        As TTI measures' effectiveness depends on daily infections, case numbers can seemingly explode when the (hard-to-estimate) TTI capacity limit is exceeded.
        Many European countries managed to stabilize case numbers over summer 2020. However, stability was short-lived, and they quickly saw case numbers rising faster than exponentially.
        \textbf{C:}
        When the external influx is low or employed strategies to reduce contacts are very effective, the stable regime can be reached. In this case, average case numbers are very low, and local outbreaks can be controlled well through local interventions. Raw data and preliminary visualizations were obtained from \cite{owidcoronavirus}.
        \textbf{D, E:} A substantial increase in Germany's effective growth rate occurred during October, suggesting that regional TTI capacity limits were exceeded. \textbf{D:} Observed case numbers were below 20 daily new cases per million until, presumably, a transition into the unstable regime took place over $\sim4$ weeks in October. The inset compares the extents of the so-called first and second waves.  \textbf{E:} Before October, the reproduction number was slightly above one (corresponding to a daily growth rate slightly above zero). More details are in Supplementary Figure~S4.}
    \label{Fig:policy_comparison}
\end{figure}

Among countries worldwide, significant variability in governmental policies and the chosen strategy to face COVID-19 significantly impact the levels of observed COVID-19 case numbers (see also \figref{Fig:policy_comparison} and Supplementary Notes~1.1).
Sustained high levels of more than 100 daily new cases per million have been observed in several (but not exclusively) American countries (\figref{Fig:policy_comparison}A). This shows that high levels of daily new infections can be maintained. However, the stringency of interventions is similar to other countries~\cite{hale_oxford_2020}, without signs of reaching population immunity.
On the other hand, very low case numbers and even local eradication have been achieved by several South- and East-Asian countries, Australia and New Zealand.
At the time of writing, these countries profit from the absorbing state at zero SARS-CoV-2 infections, but maintaining this state requires substantial international travel restrictions (\figref{Fig:policy_comparison}C).
An intermediate level of case numbers could predominantly be observed over summer 2020 in Europe.
Case numbers for many countries were typically around ten daily new cases per million (\figref{Fig:policy_comparison}B), even though contacts were only mildly restricted. These stable numbers demonstrate that also, in practice, a regime below TTI capacity limits is maintainable.
Nonetheless, in September, the spread significantly accelerated in several European countries, when case numbers began to exceed 20 to 50 daily new cases per million (Fig.~S5). Beyond these levels of case numbers, the TTI systems began to be overwhelmed, making control difficult--in line with our model's results. 

In order to focus our model on the general spreading dynamics, we made simplifying assumptions:
We assumed that spreading happens homogeneously in the population, with neither regional nor age-related differences. In reality, heterogeneous spreading can lead to regionally differing case numbers,
which illustrates the need for regional monitoring of the remaining TTI capacity to allow for early and targeted control measures.
In our scenarios, we further assumed that the population's behavior and subsequent contact reduction are constant over time (except during lockdown).
Real situations are more dynamic, necessitating frequent reevaluations of the current restrictions and mitigation measures. 
We also assumed constant TTI effectiveness if below the capacity limit. However, if case numbers are very low, all the available test- and trace efforts could be concentrated on the remaining infection chains.
This would further facilitate control at low case numbers.
Overall, our analytical results describe the general behavior across countries well and identify the relevant factors for controlling the pandemic.

Quantitatively, our assumptions regarding the efficiency of TTI are in agreement with those of other modeling studies. Agent-based models with detailed contact structures~\cite{Kucharski2020effectiveness, kerr_controlling_2020} and mean-field models~\cite{fraser2004factors,ferretti2020quantifying,Lunz2020,sturniolo2020testing} both agree that TTI measures are an essential contribution for the control of the pandemic but typically do not suffice alone. Their success strongly depends on their implementation:
Fast testing, rigorous isolation, and a large proportion of traced contacts are essential.
Given our informed assumptions about these parameters, our model shows that TTI can only compensate a basic reproduction number $R_0$ of 3.3 --- if contagious contacts are also reduced to at most \SI{61}{\percent} (95\% CI: [47,76]) compared to pre-COVID.
This is in agreement with the results of other studies~\cite{hellewell2020feasibility,ferretti2020quantifying,Davis2020ImperfectTTI,Kucharski2020effectiveness,kerr_controlling_2020}.

The capacity limit of TTI plays a central role in the control of the spread but depends strongly on the local environment.
The precise limit of TTI depends on several factors, including the number of available tests, personnel at the tracing units, potentially a tracing app~\cite{ferretti2020quantifying}, and the number of relevant contacts a person has on average.
Already the latter can easily differ by a factor of 10, depending on contact restrictions and cultural factors \cite{van2020using}.
We assumed the capacity is reached at about 85 daily new cases per million for our scenarios, which is comparably high. 
Independent of the exact value, when this limit is approached, the risk of tipping over to uncontrolled spread strongly increases, and countermeasures should be taken without delay.

Given the large deviations of the capacity limit of TTI across regions, policymakers should monitor local health authorities' and tracing agencies' capacity instead of relying on fixed limits. Health authorities can also assess whether a local outbreak is controlled or whether infection chains cannot be traced anymore, allowing an early and adaptive warning system. However, one can safely state that daily case numbers larger than 85 per million (corresponding to our modeled limit) are above the capacity limit of TTI programs in Europe, therefore requiring in any case further restrictions to reach controllable levels.  

Even given the ongoing vaccination campaigns, a low level of case numbers below TTI capacity limits remains essential. We find that during as well as after the campaigns, TTI still greatly facilitates the containment of COVID-19. The vaccinations' exact effect depends on several hard-to-model factors. It can change with newly-emerging variants of the virus, which can be more contagious, more severe, or escape the immune response. In our analysis of the effect of vaccination, we also neglected age- and high-risk-group distributions and contact networks in the population, the exact design of national vaccination plans, or the differential efficacy of vaccines against infection and severe disease. Some of these factors were taken into account in other publications \cite{bauer2021relaxing,moore2021vaccination,contreras2021risking}. However, the long-term success of vaccinations alone remains hard to predict. Thus, it is sensible to accompany mass vaccinations to achieve low case numbers in the vaccine rollout and the time beyond.

Our results show that a stable equilibrium at low case numbers can be maintained with a moderate contact reduction of about \SI{40}{\percent} less contagious contacts compared to pre-COVID-19. In terms of our parameters, this translates to a maximum --critical-- level of contacts $\RelContactsCrit$ of \SI{61}{\percent} (95\% CI: [47,76]). This level of contacts can be achieved with preventive mitigation measures, as shown by studies analyzing the effectiveness of non-pharmaceutical interventions during the first wave \cite{Brauner2020EffectivenessEurope, hsiang_effect_2020, dehning2020inferring, li_temporal_2020, sharma_understanding_2021}.
Restrictions on the maximum size of gatherings 
already lead to an effective reduction in the range of 10--\SI{40}{\percent} \cite{Brauner2020EffectivenessEurope, li_temporal_2020, dehning2020inferring, sharma_understanding_2021}.
Improved hygiene, frequent ventilation of rooms, and the compulsory use of masks can further reduce the number of infectious contacts (by a factor that is more difficult to estimate~\cite{chu_physical_2020, howard_face_2020}).
Overall, until mass vaccination plans have been deployed worldwide and available vaccines have been shown to be successful against emerging variants, the regime of low case numbers is very promising for a mid- and long-term management of the pandemic, as it poses the least burden on economy and society.

On the other hand, stabilizing the spread at higher levels of case numbers (e.g.~at the hospital capacity limit) requires more stringent and more frequent non-pharmaceutical interventions, because the TTI system cannot operate efficiently.
Examples of more stringent measures are the closure of schools and public businesses, stay-at-home orders, and contact ban policies~\cite{Brauner2020EffectivenessEurope, hsiang_effect_2020, dehning2020inferring}.

In conclusion, this paper recommends reaching and maintaining low case numbers that allow efficient TTI measures complementary to pharmaceutical interventions. 
To this end, it is mandatory to counteract local super-spreading events (or an acute influx of infections) as early as possible and to sustain a sufficient level of mitigation measures. If low case numbers are reached and maintained throughout Europe, it will be possible to lift restrictions moderately in the medium-term, and we will be better prepared for the emergence of future variants of concerns.

\section*{Methods}

\subsection*{Model overview} 
\begin{figure}[!h]
    \centering
    \includegraphics[width=10cm]{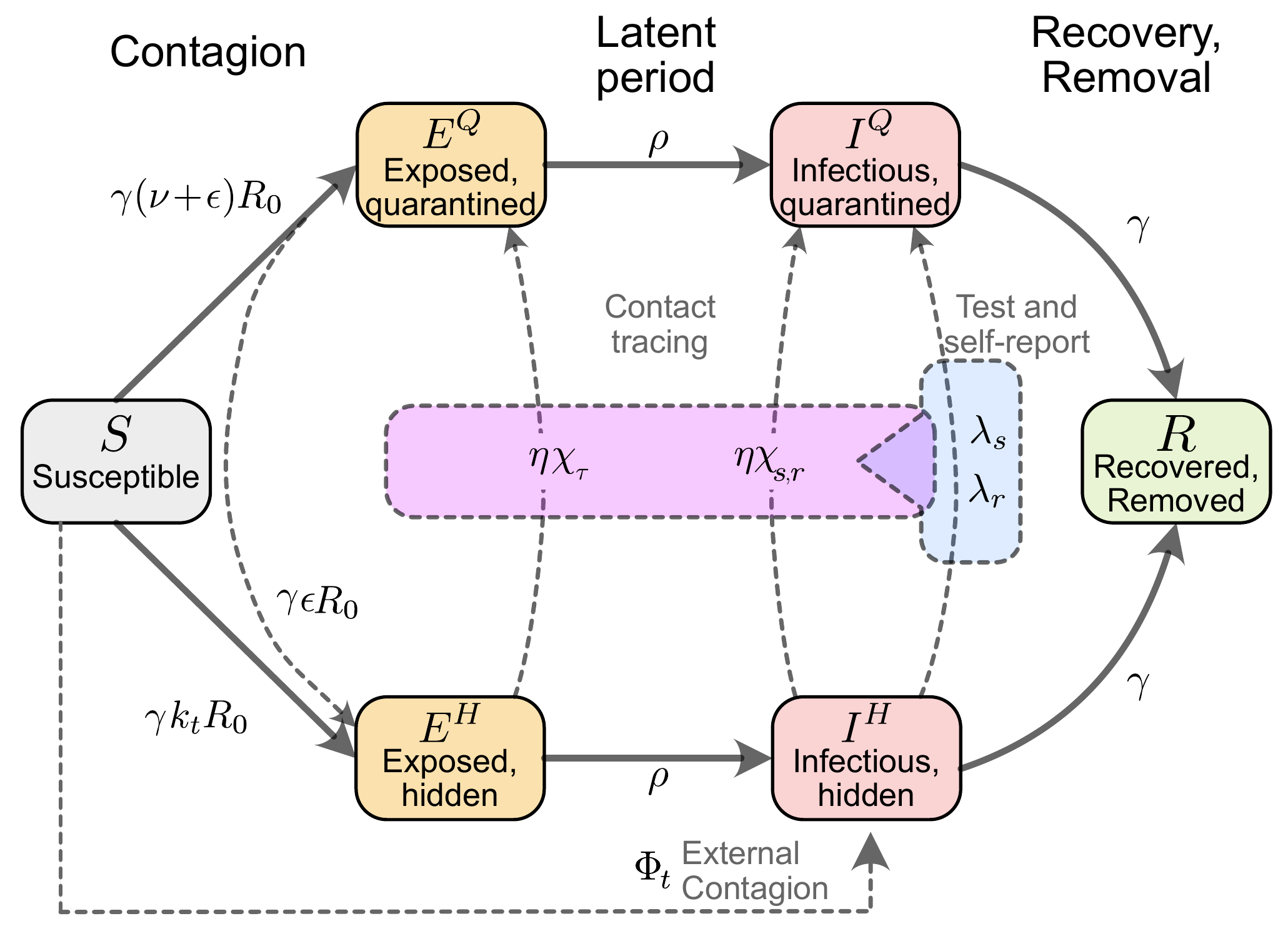}
    \caption{%
        \textbf{Flowchart of the complete model.} The solid blocks in the diagram represent different SEIR compartments for both hidden and quarantined individuals. Hidden cases are further divided into symptomatic and asymptomatic carriers (not shown).
        Solid lines represent the natural progression of the infection (contagion, latent period, and recovery). On the other hand, dashed lines account for imperfect quarantine and limited compliance, external factors, and test-trace-and-isolate policies.
        }
    \label{fig:WholeModelFlowchart}
\end{figure}

We model the spreading dynamics of SARS-CoV-2 as the sum of contributions from two pools of infectious individuals, i.e.~quarantined-isolated $\Tt$ and hidden non-isolated $\Ht$ individuals, while also modeling the infectivity timeline through the incorporation of compartments for individuals exposed to the virus ($\ET,\,\EH$), following an SEIR-like formalism. The quarantined infectious pool ($\Tt$) contains cases revealed through testing or by contact tracing and subsequently sent to quarantine/isolation to avoid further contacts as well as possible. In contrast, in the hidden infectious pool ($\Ht$), infections spread silently and only become detectable when individuals develop symptoms and get tested, via random testing in the population or as part of the chain of contacts of recently identified individuals. This second pool ($\Ht$) is called the hidden pool; individuals in this pool are assumed to exhibit the general population's behavior, thus of everyone who is not aware of being infected. Healthy individuals that can be infected belong to the susceptible pool $S$. At the same time, we assume that, after they recover and for the relatively short time frame here studied, they remain immunized in the $R$ compartment, for a graphical representation of the model, see~\figref{fig:WholeModelFlowchart}.
We model the mean-field interactions between compartments by transition rates, determining the timescales involved. These transition rates can implicitly incorporate both the disease's time course and the delays inherent to the TTI process. Individuals exposed to the virus become infectious after the latent period, modeled by the transition rate $\latRate$. We distinguish between symptomatic and asymptomatic carriers -- this is central when exploring different testing strategies (as detailed below). We also include the effects of non-compliance to TTI measures, modeled as a higher asymptomatic ratio, and imperfect contact tracing, including an explicit delay between testing and contact tracing of contacts. In the different scenarios analyzed, we include a non-zero influx $\Phit$ of new cases that acquired the virus from outside. Even though this influx makes a complete eradication of SARS-CoV-2 impossible, different outcomes in the spreading dynamics might arise depending on both contact intensity (contact level $\RelContacts$) and TTI. We then investigate the system's stability and dynamics, aiming to control the spread with a low total number of cases without necessitating a too large reduction of infectious contacts.

\subsection*{Spreading Dynamics}

Concretely, we use a modified SEIR-type model, where infected individuals can be either symptomatic or asymptomatic. They belong to hidden ($\EH,\Ht$) or a quarantined ($\ET,\Tt$) pools of infections, thus creating in total one compartment of susceptible ($S$), two compartments of exposed individuals ($\EH$, $\ET$), four compartments of infectious individuals ($\Hts$, $\Hta$, $\Tts$, $\Tta$), and one compartment for recovered/removed individuals ($R$). 

New infections are asymptomatic with a ratio $\xi$, the others are symptomatic. In all compartments, individuals are removed with a rate $\gamma$ because of recovery or death  (see~\tabref{tab:Parametros} for all parameters).  

In the hidden pools, the disease spreads according to the population's contact patterns, which can be expressed as a level $\RelContacts$ of the intensity they had before COVID-19 related contact restrictions. Defining $R_0$ as the base reproduction number without contact restrictions, the reproduction number of the hidden pool $\Ht$ is given by $\RtH$. This reproduction number reflects the disease spread in the general population without testing induced isolation of individuals. Additionally, the hidden pool receives a mobility-induced influx $\Phit$ of new infections. Cases are removed from the hidden pool (i) when detected by TTI and put into the quarantined pool $\Tt$, or (ii) due to recovery or death. 

The quarantined exposed and infectious pools $(\ET,\Tt)$ contain those infected individuals who have been tested positive as well as their positively tested contacts. Infectious individuals in $\Tt$ are (imperfectly) isolated; we assume their contacts have been reduced to a fraction $(\nu+\epsilon)$ of the ones they had in pre-COVID-19 times, of which only $\nu$ are captured by the tracing efforts of the health authorities. The subsequent infections remain quarantined, thus entering the $\ET$ pool and afterward the $\Tt$ pool. The remaining fraction of produced infections, $\epsilon$,  are missed and act as an influx to the hidden pools ($\EH$). Therefore, the overall reproduction number in the $\Tt$ pool is $\left(\nu+\epsilon\right) R_0$. 

As our model is an expanded SEIR model, it assumes post-infection immunity, which is a realistic assumption given the limited time frame considered in our analysis. Our model can also reflect innate immunity; one has to rescale the population or the reproduction number. The qualitative behavior of the dynamics is not expected to change.

\subsection*{Parameter Choices and Scenarios}\label{sec:parameters}

For any testing strategy, the fraction of infections that do not develop any symptoms across the whole infection timeline is an important parameter, and this also holds for testing strategies applied to the case of SARS-CoV-2. In our model, this parameter is called $\xiap$ and includes beside true asymptomatic infections $\xi$ also the effect of individuals that avoid testing \cite{mcdermott2020refusal}. The exact value of the fraction of asymptomatic infections $\xi$, however, is still fraught with uncertainty, and it also depends on age~\cite{Lai2020FactsMyths}. While early estimates were as high as $50\,\%$ (for example ranging from $26\,\%$ to $63\,\%$ \cite{lavezzo2020suppression}), these early estimates suffered from reporting bias, small sample sizes and sometimes included pre-symptomatic cases as well \cite{byambasuren2020estimating}. Recent studies estimate the asymptomatic transmission to be more minor \cite{cevik_virology_2020}, estimates of the fraction of asymptomatic carriers  range between $12\,\%$ \cite{byambasuren2020estimating} and $33\,\%$ \cite{pollan2020prevalence}.

Another crucial parameter for any TTI strategy is the reproduction number of the hidden infections. This parameter is by definition impossible to measure, but it is typically the main driver of the spreading dynamics. It depends mainly on the contact behavior of the population and ranges from $R_0$ in the absence of contact restrictions to values below $1$ during strict lockdown \cite{dehning2020inferring}. Herein we decided to include instead contact level compared to the pre-COVID-19 baseline $\RelContacts$ to represent the reproduction number of hidden infections $R_t^{H}=\RtH$. For the default parameters of our model, we evaluated different contact levels $\RelContacts$.

\subsection*{Testing-and-Tracing strategies}

We consider a testing-and-tracing strategy: symptom-driven testing and specific testing of traced contacts, with subsequent isolation (quarantine) of those who tested positive. Our model can also include random testing, but this case is only explored in Fig.~4 of this paper. 

\textbf{Symptom-driven testing} is defined as applying tests to individuals presenting symptoms of COVID-19. In this context, it is important to note that non-infected individuals can have symptoms similar to those of COVID-19, as many symptoms are rather unspecific. Although symptom-driven testing suffers less from imperfect specificity, it can only uncover symptomatic cases that are willing to be tested (see below). Here, \textit{symptomatic, infectious individuals} are transferred from the hidden to the traced pool at rate $\lambda_{s}$.

We define $\lambda_{s}$ as the daily rate at which symptomatic individuals get tested among the subset who are willing to get tested because of surveillance programs or self-report. As default value, we use $\lambda_{s}=0.25$, which means that, on average, an individual willing to get tested that develops COVID-19-specific symptoms would get a test within four days from the end of the latency period. Testing and isolation happen immediately in this model, but their report into the observed new daily cases $\Nhatobs$ is delayed, and so is the tracing of their contacts. 

\textbf{Tracing} contacts of positively tested infectious individuals presents a very specific test strategy and is expected to be effective in breaking the infection chains if contacts self-isolate sufficiently quickly~\cite{Kucharski2020effectiveness,Firth2020combining,kojaku2020effectiveness}. However, as every implementation of a TTI strategy is bound to be imperfect, we assume that only a fraction $\eta < 1$ of all contacts can be traced. These contacts, if tested positive, are then transferred from the hidden to the quarantined infectious pools ($\Ht\to\Tt$) with an average delay of $\tau=2$ days. The parameter $\eta$ effectively represents the fraction of secondary and tertiary infections that are found through contact tracing. As this fraction decreases when the delay between testing and contact-tracing increases, we assumed a default value of $\eta = 0.66$, i.e.~on average, only two-thirds of subsequent offspring infections are prevented. Contact tracing is mainly done manually by the health authorities in Germany. This limits the maximum number $\Nmax$ of new cases observed through testing $\ftestH$, for which contact tracing is still functional. 

\textbf{Random testing} is defined here as applying tests to individuals irrespective of their symptom status or whether they belonged to the contact chain of other infected individuals.
In our model, random testing transfers infected individuals from the hidden to the quarantined infectious pools with a fixed rate $\lambda_{r}$, irrespective of they are showing symptoms or not. In reality, random testing is often implemented as situation-based testing for a sub-group of the population, e.g.~at a hotspot, for groups at risk, or for people returning from travel. Such situation-based strategies would be more efficient than the random testing assumed in this model, which may be unfeasible at a country level due to testing limitations \cite{contreras2021challenges}. 

\subsection*{Lockdown modeling}

To assess the effectiveness of lockdowns in the broad spectrum of contact-ban governmental interventions, we model how the reduction of contacts and the duration of such restrictive regimes help lower case numbers. 

We model contact reductions as reductions in the reproduction number of the hidden population, which for these matters is presented as percentages of the basic reproduction number $R_0$, which sets the pre-COVID-19 baseline for the number of close contacts. 

For the sake of simplicity, we assume the lockdown scenarios have three stages: i) an uncontrolled regime, where the TTI capacity is overwhelmed because of high case numbers and unsustainable contact levels, reflected by a high value of $\RelContacts$ and a high influx of infections $\Phit$. ii) Lockdown is enacted, imposing a strong reduction of contacts, leading to lower values of $\RelContacts$, and borders closing, leading to a lower influx of $\Phit$. iii) Measures are relaxed, allowing higher levels of contacts $\RelContacts$ and restoring international transit. All the changes between the different regimes i$\to$ii$\to$iii are modeled as linear ramps for both parameters, which take $\Dramp = 7$ days to reach their set-point. The duration of the lockdown, namely, the time frame between the start of the restrictive measures and the beginning of their relaxation, is measured in weeks. Its default length --for analysis purposes-- is $\DL = 4$ weeks. These values have been chosen following the results of \cite{li_temporal_2020}, where the first effects of an NPI were seen after seven days, and the maximum effect after four weeks.

\subsection*{Model Equations}

The contributions of the spreading dynamics and the TTI strategies are summarized in the equations below. They govern the dynamics of case numbers between the different susceptible-exposed-infectious-recovered (SEIR) pools, both hidden (non-isolated) and quarantined. We assume a regime where most of the population is susceptible, and the time frame analyzed is short enough to assume post-infection immunity. Thus, the dynamics are completely determined by the spread (characterized by the reproduction numbers $\RtH$ and $\RtT$), the transition from exposed to infectious (at rate $\latRate$), recovery (characterized by the recovery rate $\gamma$), external influx $\Phit$ and the impact of the TTI strategies:

\begin{align}
&\frac{d S}{dt} & = & -\xunderbrace{\gamma\RtH\SM\Ht}_{\text{hidden contagion}} &-& \xunderbrace{\gamma\RtT\SM\Tt}_{\text{traced contagion}} - \xunderbrace{\SM\Phit}_{\text{ext. influx}} ~,&  \label{eq:dSdt}\\
&\frac{d \ET}{dt} & = & \xunderbrace{\gamma\nu R_0 \SM\Tt}_{\text{traced contagion}} &+& \xunderbrace{\chitau \ftrace}_{\text{contact tracing}} - \xunderbrace{\latRate\ET}_{\text{end of latency}} ~,& \label{eq:dETdt}\\
&\frac{d \EH}{dt} & = & \xunderbrace{\gamma\SM\left(\RtH\Ht+\epsilon R_0\Tt\right)}_{\text{hidden contagion}} &-& \xunderbrace{\chitau \ftrace}_{\text{contact tracing}} - \xunderbrace{\latRate\EH}_{\text{end of latency}}~,&  \label{eq:dEHdt}\\
&\frac{d \Tt}{dt} & = & \xunderbrace{\latRate\ET-\gamma\Tt}_{\text{spreading dynamics}} + \xunderbrace{\ftestH}_{\text{testing}} &+& \xunderbrace{\alphaT \ftrace}_{\text{contact tracing}} ~,&  \label{eq:dTdt}\\
&\frac{d \Ht}{dt} & = & \xunderbrace{\latRate\EH-\gamma\Ht}_{\text{spreading dynamics}} - \xunderbrace{\ftestH}_{\text{testing}} & - & \xunderbrace{\alphaT \ftrace}_{\text{contact tracing}} +\xunderbrace{\SM\Phit}_{\text{ext. influx}} ~,& \label{eq:dHdt}\\
&\frac{d \Hts}{dt} &  = & \xunderbrace{\left(1\!-\!\xi\right)\latRate\EH-\gamma\Hts}_{\text{spreading dynamics}} -\xunderbrace{\ftestHs}_{\text{testing}} & - & \left(1\!-\!\xi\right)\Bigg(\xunderbrace{\chisr\ftrace}_{\text{contact tracing}} -\xunderbrace{\SM\Phit}_{\text{ext. influx}}\Bigg)  ~,&\label{eq:dHsdt} \\
&\Hta &=&\: \Ht - \Hts,\\
&\frac{d R}{dt}    & = & \xunderbrace{\gamma\left(\Tt+\Ht\right)}_{\text{recovered/removed individuals}}.  &  & ~   & \label{eq:dRdt}
\end{align}

\subsection*{Initial conditions}

Let $x$ be the vector collecting the variables of all different pools:

\begin{equation}
    x = [S,\,\ET,\,\EH,\,\Tt,\,\Ht,\,\Hts,\,R].
\end{equation}

We assume a population size of $M=10^6$ individuals, so that $\sum_{i\neq 6} x_i = M$, and a prevalence of $I_0 = 200$ infections per million, so that $\Tt(0) = I_0$. Assuming that the hidden amount of infections is in the same order of magnitude $I_0$, we would have $\Ht(0)=I_0,\,\Hts(0) = (1-\xi)I_0$. We would expect the exposed individuals to scale with $\RtH I_0$, but we rather assume them to have the same size of the corresponding infectious pool. To calculate the initially susceptible individuals, we use $S(0) = 1-\sum_{i\neq\{1,6\}} x_i$.

\subsection*{Effect of delays and capacity limit on the effectiveness of TTI strategies}

In this section we discuss further details on the derivation of the different parameters and variables involved in equations \eqref{eq:dSdt}--\eqref{eq:dRdt}. First, as we assume contact tracing to be effective after a delay of $\tau$ days, some of the individuals who acquired the infection from those recently tested might have also become infectious by the time of tracing. Moreover, a fraction of those who became infectious might also have been tested by the tracing time, should they have developed symptoms. 

Furthermore, we give explicit forms for $\ftestH$ and $\ftrace$ the number of cases identified respectively by testing and contact-tracing. When surpassing TTI capacity, we assume that both testing and contact-tracing change their dynamics simultaneously. This happens when the daily amount of cases identified by testing $\ftestH$ overpasses the TTI threshold $\Nmax$. After being overwhelmed, the overhead testing would change its rate $\lambda_s\to\lambda_s'$, as only patients with a more specific set of symptoms would be tested. Nonetheless, the contact-tracing efforts can only follow the contacts of those $\Nmax$ observed cases, identifying a fraction $\eta$ of the offspring infections they produced in their infectious period spent in unawareness of their state. The possibility of random testing is analyzed in Supplementary Note~1.6.

\subsubsection*{Limited testing capacity leading to lower testing rates}

In the first stages of an outbreak, individuals with any symptoms from the broad spectrum of COVID-19-related symptoms would be tested, disregarding how specific those symptoms are. 
At this stage, we assume that the rate  at which symptomatic individuals are tested is $\lambda_s$, such that the number of individuals identified through testing (which, for simplicity, is assumed to be solely symptom-driven, i.e., $\lambda_r=0$) is given by

\begin{equation}
    \ftestH = \ftestHs  = \lambda_s \Hts.
\end{equation}
If in addition some random testing, independent on symptomatic status, is performed ($\lambda_r\neq0$), then $\ftestH \neq \ftestHs$. For this case see S1.6.

When reaching the daily number $\Nmax$ of positive tests, the testing capacity is reached. We then assume that further tests are only carried out for a more specific set of symptoms, leading to a smaller fraction of the tested population. We, therefore, implement the testing capacity as a soft threshold. Assuming that after reaching $\Nmax$, the testing rate for further cases would decrease to $\lambda_s'$, the testing term $\ftestH$ would be given by

\begin{equation}
    \ftestH  =   \lambda_s\min\left(\Hts,\Htsmax\right) + \lambda_s'\max\left(0,\Hts\!-\!\Htsmax\right),
\end{equation}

where $\Htsmax$ represent the size of the infectious -symptomatic, hidden- pool i.e, $\Htsmax = \frac{\Nmax}{\lambda_s}$.

\subsubsection*{Modeling the number of traced individuals}

To calculate the number of traced individuals, we assume that a fraction $\eta$ of the newly tested individuals' contacts, and therefore their offspring infections, will be traced and subsequently quarantined.  
However, in the presence of TTI, individuals stay on average a shorter amount of time in the infectious pool because they are quarantined before recovering. Therefore, the number of offspring infections has to be corrected by a factor, the average residence time in the infectious pool. For the case $\lambda_r=0$, the average residence time is $\frac{1}{\gamma + \lambda_s}$, as $\Hts$ is emptied by $-\gamma\Hts-\ftestHs=-(\gamma+\lambda_s)\Hts$, i.e., with a rate $\gamma+\lambda_s$. The average residence time in the absence of TTI (natural progression of the disease) is $\frac{1}{\gamma}$. Dividing these two times gives us the wanted correction factor.
Thus, the number of traced persons $\ftrace$ at time $t$ is a fraction $\eta$ from the offspring infections generated during the residence time, per each individual:

\begin{equation}
    \ftrace \left(t\right)=  \eta R_{t-\tau}  \frac{\gamma}{\gamma + \lambda_s} \ftestH\left(t-\tau\right),
\end{equation}

where $R_{t-\tau}$ represents the effective reproduction number:
\begin{equation}
R_{t-\tau}=k_{t-\tau}R_0\frac{S}{M}.
\end{equation}

In other words, the number of infectious individuals found by contact tracing at time $t$ are a fraction $\eta$ of the number of offspring infections generated by individual while they were untested $R_{t-\tau}\frac{\gamma}{\gamma + \lambda_s}$, times the number of individuals tested $\tau$ days ago $\ftestH\left(t-\tau\right)$. However, when the TTI capacity is overwhelmed, we assume that the number of traced individuals is limited, that only the contacts of $\Nmax$ individuals (already introduced in the previous section) can be traced:

\begin{equation}
    \ftrace \left(t\right)=  \eta R_{t-\tau}  \frac{\gamma}{\gamma + \lambda_s} \Nmax.
\end{equation}

\subsubsection*{Individuals becoming infectious or being tested by the time of tracing}

The traced individuals are removed from either the exposed hidden pool $\EH$ or from the infectious hidden pool $\Ht$ after a delay of $\tau$ days after testing. As we assume a tracing delay $\tau$ of only two days, a significant fraction of the traced individuals would still be in exposed compartments by the time of contact tracing. However, some might already become infectious by that time. 
To calculate the exact fraction of individuals remaining in the hidden exposed pool by the time of tracing, we proceed as follows. Let $s \in I_\tau = \left[0,\,\tau\right]$ be the time elapsed from the moment of testing. The emptying of the normalized exposed compartment (denoted $\widetilde{\EH}$) due to progression to the infectious stage follows  first-order kinetics:
\begin{equation}\label{eq:dXds}
    \frac{d \widetilde{\EH}}{ds} = -\latRate \widetilde{\EH},\qquad \widetilde{\EH}(0) = 1
\end{equation}

The solution of~\eqref{eq:dXds} is given by $\widetilde{\EH(s)} = \exp\left(-\latRate s\right)$. Therefore, we define $\chitau$ as the fraction of the traced individuals remaining in the $\EH$ compartment at $s=\tau$:

\begin{equation}
\chitau = \exp\left(-\dis\latRate\tau\right)\label{eq:chitau}.
\end{equation}

The remaining individuals are removed from the infectious compartment, which are then simply described by the fraction
\begin{align}
    \chir     = 1-\chitau.\label{eq:chir}
\end{align}

This, however, only holds for the asymptomatic hidden infectious pool. For the symptomatic hidden pool $\Hts$, we do not want to remove the individuals who have already been tested, as they would be removed twice. For modeling the fraction of non-tested individuals remaining in the normalized symptomatic infectious compartment (denoted $\widetilde{\Hts})$, we couple two first-order kinetics:

\begin{equation}\label{eq:dYds}
    \frac{d \widetilde{\Hts}}{ds} =  -\lambda_s\, \widetilde{\Hts} + \rho \widetilde{\EH},\qquad \widetilde{\Hts}(0) = 0.
\end{equation}

The solution of~\eqref{eq:dYds} depends on whether  $ \lambda_s = \latRate$ or not. The solution at $s=\tau$ which is the fraction of traced individuals removed from $\Hts$ is given by:

\begin{align}
    \chisr  & = \left\{ \begin{array}{ll} \dis\latRate\tau\exp\left(-\dis\latRate\tau\right) & \text{if }  \lambda_s\approx\latRate, \\                       \dis\frac{\latRate}{\lambda_s-\latRate}\left(\exp\left(-\dis\latRate\tau\right)-\exp\left(-\dis\lambda_s\tau\right)\right)&                   \text{else.}\end{array} \right. \label{eq:chisr}
\end{align}

For the case $\lambda_r\neq 0$, the reader is referred to the Supplementary Note~1.6.

\subsection*{Including the effects of ongoing vaccination campaigns}

To incorporate the effects of COVID-19 vaccination programs in our model, we made some simplifying assumptions. First, we assume that vaccinated individuals have a probability $\kappa$ of not being infected even if they have a contact with somebody infectious, then not contributing to the spreading dynamics. We define this parameter as the "vaccine efficacy against infection", which has been reported around 50--90\% for available vaccines \cite{thompson2021interim,dagan2021BNT,voysey2021safety}.  Thus, they can be assumed to have developed perfect immunity and therefore can be removed from the susceptible ($S$) and put into the removed compartment ($R$). The assumption above also implies that these individuals would not take part in the TTI scheme, which would resemble the growing trend of "vaccination passports". Second, we assume a logistically increasing daily vaccination rate $v(t)$ consistent with the projections in the European Union (EU) and assume that 70\% of the total population get vaccinated (see \figref{fig:vaccination}~B, E). This would, e.g., in Germany, amount to roughly \SI{80}{\%} of the adult population (16+ years old) accepting the offer of vaccination, since as of now, none of the available vaccines is approved for children. We find that efficient TTI can significantly enhance the effect of the growing immunity.

To include gradually growing immunity due to an ongoing vaccination campaign we modify \eqref{eq:dSdt} with an additional term $- \kappa\cdot v(t)$, as well as \eqref{eq:dRdt} with a $+ \kappa\cdot v(t)$, with a daily vaccination rate 

\begin{equation}
    v(t) =  \frac{\num{9.3e3}}{1+\exp\left(-0.025(t-150)\right)}\,\text{doses per million per day},
\end{equation}
centered at $t=0$, which denotes the start of the vaccinations. This logistic increase in vaccination rates and parameters involved were adapted from \cite{bauer2021relaxing} and roughly mirrors the projected vaccinations in the EU (projections dated to the beginning of February 2021). The factor $\num{9.3e3}$ is determined assuming that after $t_{\rm ref}=220$ days, 70\% of the population would be vaccinated. Using as reference the age distribution of Germany, the above would amount to roughly a \SI{80}{\%} of the adult population (16+ years old) accepting the offer of vaccination. After that time, for simplicity, we assume the vaccination stops (and therefore $v(t)=0$, for $t>220$).

This treatment of the vaccination is simplistic. In reality, most currently available vaccines imply receiving two doses in the span of a few weeks, where the first only gives partial protection. Furthermore, vaccinated individuals need some time to develop a proper immune response after receiving the vaccine \cite{polack2020safety}, in which they can still get infected. We also do not incorporate the efficacy of vaccines against a severe course of the disease or death. Since vaccinated but yet infected individuals would have a lower chance of being admitted to the hospital, this would falsify our assumption that hospital capacity can be adequately measured by case numbers alone. As more and more of the daily new infections correspond to individuals already vaccinated, hospitals would only fill up at higher case numbers. To include this effect, the distribution of high-risk groups in the population and the prioritized vaccination programs would have to be taken into account. Including all this is beyond the scope of this work. We addressed in a separate work, building on the results presented herein \cite{bauer2021relaxing}. Yet, this simplified implementation is sufficient for our qualitative assessments.

\subsection*{Central epidemiological parameters that can be observed}

In the real world, the disease spread can only be observed through testing and contact tracing. While the \textit{true} number of daily infections $N$ is a sum of all new infections in the hidden and traced pools, the \textit{observed} number of daily infections $\Nhatobs$ is the number of new infections discovered by testing, tracing, and surveillance of the contacts of those individuals in the quarantined infectious pool $\Tt$, delayed by a variable reporting time. This includes internal contributions and contributions from testing and tracing:

\begin{align}
    N &= \xunderbrace{\gamma\RtH\SM\Ht}_{\text{hidden contagion}} + \xunderbrace{\gamma \left(\nu+\epsilon\right)R_0\SM\Tt}_{\text{traced contagion}}+ \xunderbrace{\SM\Phit}_{\text{ext. influx}}
    \label{eq:N}\\
    \Nhatobs &=  \Big[\xunderbrace{\latRate\ET}_{\text{traced contagion}}+\xunderbrace{\ftestH+ \alphaT\ftrace}_{\text{TTI}}\,\Big]  \circledast \mathcal{K},
    \label{eq:Nreport}
\end{align}
where $\circledast$ denotes a convolution and $\mathcal{K}$ an empirical probability mass function that models a variable reporting delay, inferred from German data (as the RKI reports the date the test is performed, the delay until the appearance in the database can be inferred): The total delay between testing and reporting a test corresponds to one day more than the expected time the laboratory takes for obtaining results, which is defined as follows: from testing, \SI{50}{\%} of the samples would be reported the next day, \SI{30}{\%} the second day, \SI{10}{\%} the third day, and further delays complete the remaining \SI{10}{\%}, which for simplicity we will truncate at day four. Considering the extra day needed for reporting, the probability mass function for days 0 to 5 would be given by $\mathcal{K}=[0,\,0,\,0.5,\,0.3,\,0.1,\,0.1]$. The spreading dynamics are usually characterized by the observed reproduction number $\Rtobs$, which is calculated from the observed number of new cases $\Nhatobs(t)$. We here use the definition underlying the estimates that Robert-Koch-Institute publishes, the official body responsible for epidemiological control in Germany~\cite{anderHeiden2020Schatzung}: the reproduction number is the relative change of daily new cases $N$ separated by four days (the assumed serial interval of COVID-19):

\begin{equation}
    \Rtobs = \frac{\Nhatobs(t)}{\Nhatobs(t-4)}
\end{equation}

In contrast to the original definition of $\Rtobs$~\cite{anderHeiden2020Schatzung}, we do not need to remove real-world noise effects by smoothing this ratio.

\subsection*{Numerical calculation of solutions and critical values.}

The numerical solution of the delay differential equations governing our model were obtained using a versatile solver that tracks discontinuities and integrates with the explicit Runge-Kutta (2,3) pair, \texttt{@dde23} implemented in MATLAB (version 2020a), with default settings. This algorithm allows the solution of non-stiff systems of differential equations in the shape $y'(t)=f(t,y(t),y(t-\tau_1),...,y(t-\tau_k)$, for a set of discrete lags $\{\tau_i\}_{i=1}^{k}$. Suitability and details on the algorithm are further discussed in \cite{shampine2001solving}. 

To derive the tipping point between controlled and uncontrolled outbreaks (e.g., critical, minimal required contact reduction $\RelContactsCrit$ for both stability and metastability), and to plot the stability diagrams, we used the \texttt{@fzero} MATLAB function, and the linear approximation of the system of DDE \eqref{eq:dETdt}--\eqref{eq:dHsdt} for the $\SM\approx 1$ limit. This function uses a combination of bisection, secant, and inverse quadratic interpolation methods to find the roots of a function. For instance, following the discussion of Supplementary Notes~1.2, the different critical values for the contact reduction $\RelContactsCrit$ were determined by systematically solving the nonlinear eigenvalues problem for stability \cite{jarlebring2008some}, where the solution operation was approximated with a Chebyshev differentiation matrix \cite{trefethen2000spectral}.

We also study the effect of dividing the exposed compartment into three sub-compartments, thereby reducing the variability of the latent period distribution (understood as the distribution of waiting times from being infected until becoming infectious). We explored this extended system's linear stability in Supplementary Note~1.8 and confirmed that using a single compartment efficiently characterizes the tipping points.

\begin{table}[htp]\caption{Model parameters.}
\label{tab:Parametros}
\centering
\begin{tabular}{l p{6cm} lll p{3cm}}\toprule
Parameter       & Meaning                       & \makecell[l]{Value \\ (default)}    & \makecell[l]{Range\\ }         & Units             &   Source  \\\midrule
$M$             & Population size               & $\num{1000000}$       &               & people          &   -       \\
$R_0$           & Basic reproduction number     & 3.3                   & 2.2--4.4        & \SI{}{-}  & \cite{ZHAO2020214,alimohamadi2020151}\\
$\nu$           & Registered contacts (quarantined)& 0.075      &        & \SI{}{-}          &  Assumed   \\
$\epsilon$      & Lost contacts (quarantined)     & 0.05      & & \SI{}{-}          &  Assumed    \\
$\gamma$        & Recovery/removal rate         & 0.10      & 0.08--0.12& \SI{}{day^{-1}}  &  \cite{he2020temporal,pan2020time}        \\
$\xi$           & Asymptomatic ratio            & 0.32      & 0.15--0.43     & \SI{}{-}          &   \cite{byambasuren2020estimating,pollan2020prevalence,lavezzo2020suppression}       \\
$\lambda_s$     & Symptom-driven testing rate   & 0.25      & 0--1 & \SI{}{day^{-1}}          & Assumed \\
$\lambda_s'$    & Symptom-driven testing rate (reduced capacity)   & 0.1      &  & \SI{}{day^{-1}}          & Assumed \\
$\lambda_r$     & Random testing rate & 0.0      & 0.0--0.1 & \SI{}{day^{-1}}          & Assumed \\
$\eta$          & Tracing efficiency            & 0.66      &      & \SI{}{-}          &  Assumed  \\
$\tau$          & Contact tracing delay         & 2     &           & \SI{}{days}           &  Assumed  \\
$\Nmax$         & Maximal tracing capacity      & 50        & 10--75  & \SI{}{cases\, day^{-1}}          &  Assumed  \\
$\Phit$         & External influx               & 1           & & \SI{}{cases\, day^{-1}}        &  Assumed  \\
$\latRate$      & Exposed-to-infectious rate    & 0.25     &           & \SI{}{day^{-1}}       &   \cite{bar2020science,li2020substantial}      \\
$\DL$           & lockdown duration             & 4     &   0-8        & \SI{}{weeks}           &   \cite{li_temporal_2020}    \\
$\Dramp$        & Phase-transition duration (lockdown) & 1     &           & \SI{}{weeks}       &   \cite{li_temporal_2020}    \\\midrule
$\chitau$   & Fraction of contacts traced before becoming infectious & 0.61    &          & \SI{}{-}        &  eq~\eqref{eq:chitau}    \\
$\chisr$   & Fraction of contacts traced after becoming infectious, before being tested (symptomatic and random) & 0.30    &          & \SI{}{-}      &  eq~\eqref{eq:chisr}    \\
$\chir$   & Fraction of contacts traced after becoming infectious, before being tested (random) & 0.39    &          & \SI{}{-}       &  eq~\eqref{eq:chir}    \\\bottomrule
\end{tabular}%
\end{table}

\begin{table}[htp]\caption{Model variables.}
\label{tab:Variables}
\centering
\begin{tabular}{l p{4cm} l  p{9cm} }\toprule
Variable & Meaning & Units & Explanation\\\midrule
$S$ & Susceptible pool & \SI{}{people} & non-infected people that may acquire the virus.  \\
$\ET$ & Exposed pool (quarantined)   & \SI{}{people} & Total quarantined exposed people. \\
$\EH$ & Exposed pool (hidden)    & \SI{}{people} & Total non-traced, non-quarantined exposed people.\\
$\Hts$ & Infectious pool (hidden, symptomatic) & \SI{}{people} & Non-traced, non-quarantined people who are symptomatic.\\
$\Ht$ & Infectious pool (hidden) & \SI{}{people} & Total non-traced, non-quarantined infectious people.\\
$\Tt$ & Infectious pool (quarantined) & \SI{}{people} & Total quarantined infectious people.\\
$N$ & New infections (Total) & \SI{}{cases\, day^{-1}} & Given by:  $N = \gamma\RtH\Ht +\gamma \left(\nu+\epsilon\right) R_0\Tt+\SM\Phit$.\\
$\RelContacts$ & Contact reduction & \SI{}{\%} & Reduction of infectious contacts, related to pre-COVID-19 times. \\
$\Nhatobs$ & Observed new infections (influx to traced pool) &  \SI{}{cases\, day^{-1}} & Daily new cases, observed from the quarantined pool; delayed because of imperfect reporting and realistic contact tracing. \\
$\Rtobs$ & Observed reproduction number&  \SI{}{-} & The reproduction number that can be estimated only from the observed cases: $\Rtobs = \Nhatobs(t)/\Nhatobs(t-4)$. \\
$\ftestH$ & Number of cases found through testing&  \SI{}{people} & Cases can be found either through symptomatic or random testing $\ftestH = N^{\text{test}}_r + \ftestHs$. \\
$\ftrace$ & Number of uncovered infections through tracing&  \SI{}{people} & This number is limited (depending on the reproduction number) by the maximal tracing capacity $\Nmax$ \\
\bottomrule
\end{tabular}%
\end{table}

\newpage


\section*{Author Contributions}

Conceptualization: S.C., J.D., V.P.\\
Methodology: S.C., J.D., V.P.\\
Software: S.C., S.B.M.\\
Validation: S.C., J.D., S.B.M., S.B., V.P.\\
Formal analysis: S.C., J.D., S.B.M., P.S., S.B., V.P.\\
Investigation: S.C., J.D., S.B.M., P.S., S.B.\\	
Writing - Original Draft: S.C., J.D., S.B.M., P.S., V.P. \\	
Writing - Review \& Editing: S.C., J.D., S.B.M., P.S., S.B., V.P.\\
Visualization: S.C., P.S., S.B.M.\\	
Supervision: V.P. \\	

\section*{Competing Interests}
V.P. is currently active in various groups to advise the government. 

\section*{Acknowledgements} 
We thank Melanie Brinkmann  and Álvaro Olivera-Nappa for their helpful comments and encouraging feedback.
We thank M. Loidolt, Michael Wibral, Johannes Zierenberg, Joel Wagner, and Anamaria Sanchez for carefully reading, commenting and improving the manuscript. We thank the Priesemann group for exciting discussions and for their valuable input. 
\textbf{Funding:} All authors received support from the Max-Planck-Society (Max-Planck-Gesellschaft MPRG-Priesemann). S.C. acknowledges funding by the Centre for Biotechnology and Bioengineering - CeBiB (PIA project FB0001, ANID, Chile). J.D. and P.S. acknowledge funding by SMARTSTART, the joint training program in computational neuroscience by the VolkswagenStiftung and the Bernstein Network.  S.B.M and S.B were financially supported by the German Federal Ministry of Education and Research (BMBF) as part of the Network University Medicine (NUM), project egePan, funding code: 01KX2021.
\subsection*{Data and Materials Availability}
All data needed to evaluate the conclusions in the paper are present in the paper and/or the Supplementary Materials. Raw data and preliminary visualizations of Fig.~6, ~S4, and~S5 were obtained from \cite{owidcoronavirus}. Analysis code available in \url{https://github.com/Priesemann-Group/covid19_metastability} and permanently stored in \cite{zenodo}. Additionally, an interactive platform to simulate scenarios different from those presented here \url{http://covid19-metastability.ds.mpg.de/}.

\section*{List of Supplementary Materials}

\begin{itemize}
    \item \textbf{Supplementary Note S1.1:} Strategies to face COVID-19 differ among countries.
    \item \textbf{Supplementary Note S1.2:} Linear stability analysis and uncertainty propagation.
    \item \textbf{Supplementary Note S1.3:} On the contact reduction required for achieving early population immunity.
    \item \textbf{Supplementary Note S1.4:} Calculating the increase in the level of contacts allowed with increased immunity.
    \item \textbf{Supplementary Note S1.5:} Inferring the reproduction number of COVID-19 from Jun to Oct. 2020.
    \item \textbf{Supplementary Note S1.6:} On the incorporation of random testing in the TTI scheme.
    \item \textbf{Supplementary Note S1.7:} More analytical insights into the TTI-based metastable regime.
    \item \textbf{Supplementary Note S1.8:} Exploring the effect of more compartments for the exposed individuals.
\end{itemize}

\newpage
\renewcommand{\thefigure}{S\arabic{figure}}
\renewcommand{\figurename}{Supplementary~Figure}
\setcounter{figure}{0}
\renewcommand{\thetable}{S\arabic{table}}
\renewcommand{\tablename}{Supplementary~Table}

\setcounter{table}{0}
\renewcommand{\theequation}{\arabic{equation}}
\setcounter{equation}{0}
\renewcommand{\thesection}{S\arabic{section}}
\setcounter{section}{0}
\setcounter{page}{1}
\section{Supplementary Notes of "Low case numbers enable long-term stable pandemic control without lockdowns"}

\begin{figure}[th!]
    \centering
    \includegraphics[width=16cm]{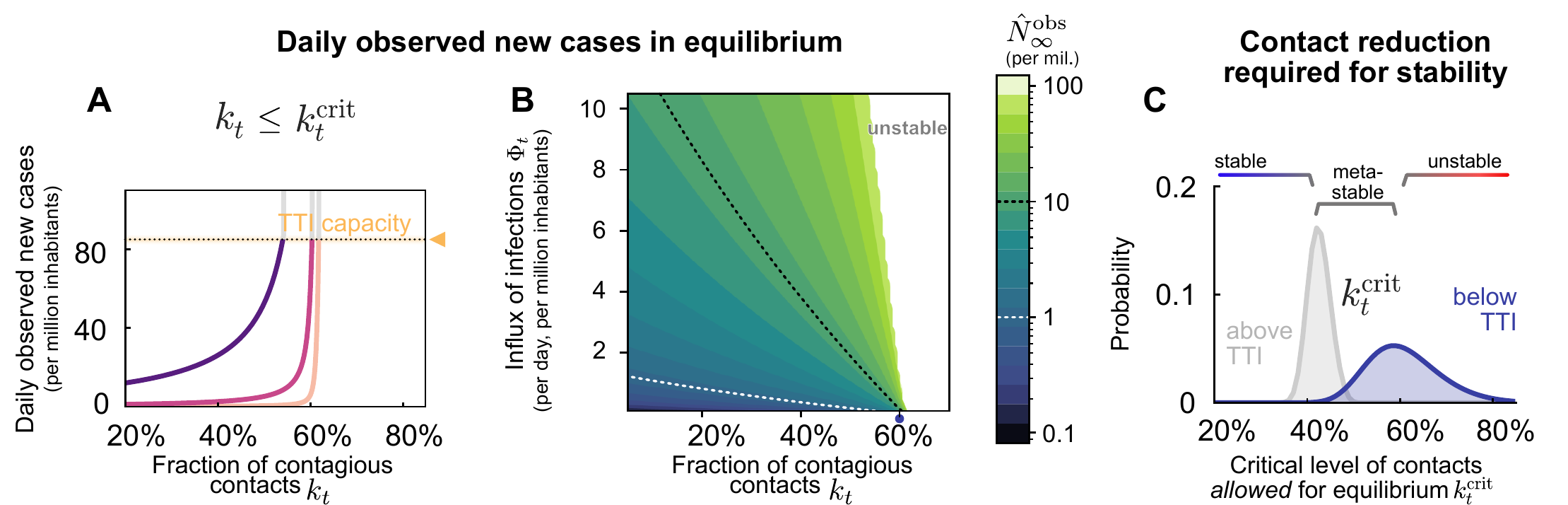}
    \caption{\textbf{In the stable and metastable regimes, daily new cases approach an equilibrium value $\Nequil$ that depends on the level of contacts $\RelContacts$ and an external influx of new cases $\Phit$.}
    \textbf{A:} 
    The equilibrium value $\Nequil$ increases steeply with higher $\RelContacts$ and then destabilizes when surpassing a critical value $\RelContactsCrit$. Thus, contact reduction has to be sufficiently strong to keep the case numbers within TTI capacity.
    A certain degree of external influx $\Phit$ can be compensated, but in general, $\Phit$ can put stability at risk. $\Phit=1$ one daily new case per million is our default parameter for the influx.
    \textbf{B:}
    The $\Nequil$ is below 10 for a large fraction of combinations of $\RelContacts$ and $\Phit$, thus well below the capacity limit of TTI. 
    \textbf{C:}
    The critical value $\RelContactsCrit$, which marks the transition between stable and (meta)stable spread; thus  $\RelContactsCrit$ displays the maximal level of contacts that is allowed while controlling the spread (and stabilize case numbers).
    If case numbers are below the TTI capacity limit, a contact level of at most $\RelContactsCrit=\SI{61}{\%}$ is allowed for stabilization (blue). 
    If case numbers, however, are above the TTI limit, stronger contact reduction is necessary for stabilization, thus allowing a lower level of contacts ($\RelContactsCrit=\SI{42}{\%}$, gray).
    Confidence intervals originate from error propagation of the uncertainty of the underlying model parameters. All model parameters are listed in \tabref{tab:Parameters_uncertainty} and the full uncertainty analysis is in \figref{fig:UncertaintyPropagation}.
    }\label{extfig:N_eq}
\end{figure}

\subsection{Strategies to face COVID-19 differ among countries.}\label{sec:worldpicture}

Several South- and East-Asian countries and Australia, and New Zealand have achieved very low case numbers and even local eradication. These countries reached very low values below one daily new case per million (median $0.5$, \figref{Fig:policy_comparison}~C). If local eradication is successful, these countries can profit from the absorbing state of zero SARS-CoV-2 infections, i.e.~after local eradication, new infection chains are only started if a virus is \textit{de novo} carried into the country~\cite{Siegenfeld2020,bittihn_containment_2020}. However, the local eradication is constantly put at risk by the undetected influx of new viruses from abroad, requiring rigorous quarantine for international travel, and -- once the spread got out of control -- decisive action to completely stop all infection chains. However, the more countries adhere to this strategy successfully, the closer one may get to global eradication.

In many European countries over the summer, case numbers were relatively low, typically around ten daily new cases per million (\figref{Fig:policy_comparison}~B). During that time, contacts were only mildly or moderately restricted, and containment was complemented by hygiene, masks, and other preventive measures. 
However, in summer and autumn, most European countries developed a second wave \figref{extfig:europe_bayesian}. The causes are undoubtedly diverse, from increasing contact rates to seasonal effects and travel-related influx. Seasonal effects alone cannot explain the second wave, as neighboring countries like Portugal versus Spain or Finland versus Sweden show remarkably different dynamics (see~\cite{owidcoronavirus}). Hence, it seems to be possible to maintain an equilibrium at relatively low case numbers. However, that equilibrium is fragile at high case numbers, and novel waves can emerge at any time.

Sustained high levels of case numbers have been observed in several countries such that TTI probably could not be performed effectively. Around 130 daily new cases per million have been observed, e.g.~in many (but not all) American countries (median $129.54$, \figref{Fig:policy_comparison}~A). It shows that high levels of daily new infections can be maintained in principle. However the stringency of interventions is similar or higher compared to other countries (see~\cite{hale_oxford_2020}), and even with these high numbers it will probably take about $\num{200000} / 150 = 1333$ to $\num{700000}/150 = 4666$ days, thus several years, until 20 to \SI{70}{\percent} of the population is infected and population immunity reached -- assuming the duration of individual immunity is long enough. That high level of new infections leads to a considerable death toll, as currently about \SI{1.5}{\percent} of the infected individuals would die (depending on age structure~\cite{Linden2020DAE,Levin2020}). Moreover, containment measures like quarantine become unsustainable because, if implemented, each one of the 200 daily new infected cases would require the quarantine of 5-50 people (their high-risk contacts) for about ten days, causing 1 - \SI{10}{\percent} of the population being in quarantine at any given day. Therefore the alleged economic and social benefits of such a strategy\cite{Alwan2020,Priesemann2020panEur} may be questionable.

\subsection{Linear stability analysis and uncertainty propagation}\label{sec:linearStab}

For analyzing the stability of the governing differential equations, namely, whether an outbreak could be controlled, we studied the linear stability of the system. Moreover, we consider that, within the time-frame considered for stability purposes, the fraction $\SM$ would remain somewhat constant, we consider the linearized version of equations~\eqref{eq:dETdt}--~\eqref{eq:dHsdt}, defining a system of delay differential equations for the variables $x(t) = \left[\ET(t);\,\EH(t);\,\Tt(t);\,\Ht(t);\,\Hts(t)\right]$. We define matrices $A$ and $B$ as:

\begin{eqnarray}
    A & = & 
    \begin{pmatrix} 
    -\latRate   & 0                                     & \nu\gamma R_0         & 0                         & 0 \\
    0           & -\latRate                             & \epsilon\gamma R_0    & \gamma\RtH    & 0 \\
    \latRate    & 0                                     & -\gamma               & \lambda_r                 & \lambda_s \\
    0           & \latRate                              & 0                     & -\gamma-\lambda_r         & -\lambda_s\\
    0           & \left(\!1-\!\xi\!\right)\latRate      & 0                     &        0                  & -\gamma -\lambda_r -\lambda_s \\
    \end{pmatrix}\\
    B & = & 
    \begin{pmatrix} 
    0   & 0 & 0 & \lambda_r^{\rm eff} \chitau   &\lambda_s^{\rm eff} \chitau    \\
    0   & 0 & 0 & -\lambda_r^{\rm eff} \chitau  & -\lambda_s^{\rm eff} \chitau  \\
    0   & 0 & 0 & \lambda_r^{\rm eff}  \left(\xi\chir + \left(\!1-\!\xi\!\right)\chisr\right)  & \lambda_s^{\rm eff}  \left(\xi\chir + \left(\!1-\!\xi\!\right)\chisr\right)               \\
    0   & 0 & 0 & -\lambda_r^{\rm eff} \left(\xi\chir + \left(\!1-\!\xi\!\right)\chisr\right) & -\lambda_s^{\rm eff} \left(\xi\chir + \left(\!1-\!\xi\!\right)\chisr\right)\\
    0   & 0 & 0 & -\lambda_r^{\rm eff} \left(\!1-\!\xi\!\right)\chisr                      &   -\lambda_s^{\rm eff} \left(\!1-\!\xi\!\right)\chisr\\
    \end{pmatrix}\eta\RelContacts R_0,
\end{eqnarray}

where 

\begin{equation}
    \lambda_r^{\rm eff} = \frac{\gamma\lambda_r}{\lambda_r+\gamma},\qquad \lambda_s^{\rm eff} = \gamma\left(\frac{\lambda_s+\lambda_r}{\gamma+\lambda_s+\lambda_r}-\frac{\lambda_r}{\lambda_r+\gamma}\right).
\end{equation}

The equations governing the dynamics for vector $x(t)$ are then presented in their matrix form:

\begin{equation}
    x'(t) = Ax(t)+Bx(t-\tau).
\end{equation}

We determine the maximum --critical-- level of contacts $\RelContactsCrit$, for which exponential solutions would be asymptotically stable. Eigenvalues were determined by systematically solving the nonlinear eigenvalues problem for stability \cite{jarlebring2008some}, where the solution operation was approximated with a Chebyshev differentiation matrix \cite{trefethen2000spectral}. Eigenvalues, in this sense, would be solutions of the scalar equation

\begin{equation}\label{eq:nonlinearEig}
    \det\left(-sI+A+e^{-s\tau}B\right)=0
\end{equation}

Noting that $A$ and $B$ explicitly depend on the model parameters, we numerically explore which combinations would result in stable, metastable, or unstable case numbers. Concretely, we studied the maximum --critical-- level of contacts allowed for stability $\RelContactsCrit$ in two scenarios; i) low case numbers, and TTI fully operative (both testing and contact tracing), and ii) high case numbers, above the TTI limit, where testing would be inefficient and solely symptom driven ($\lambda_s=\lambda_s',\,\eta=\lambda_r=\lambda_r'=0$).

\begin{figure}[!h]
    \centering
    \includegraphics[width=15cm]{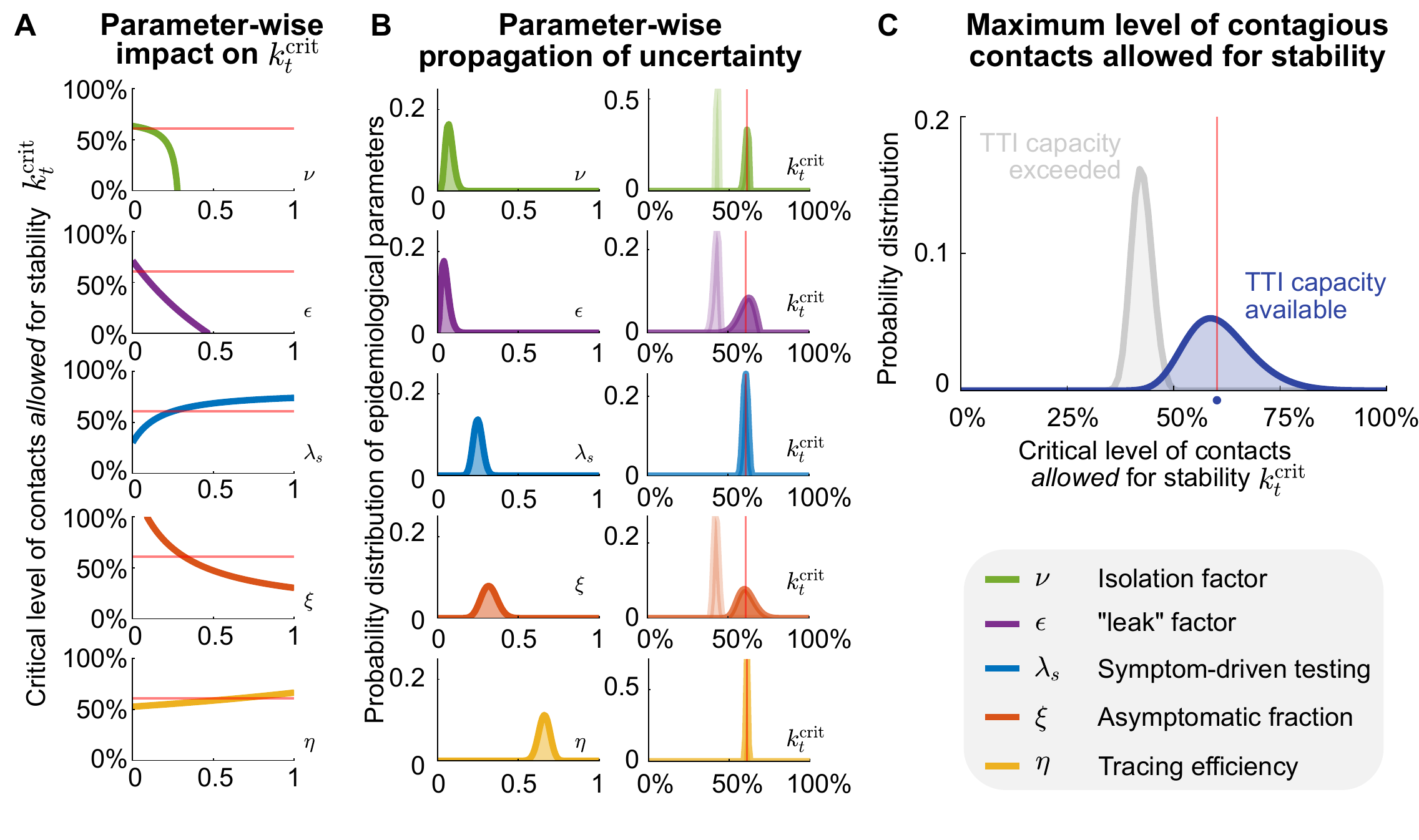}
    \caption{%
        \textbf{Propagation of TTI-parameter uncertainties to the critical level of contacts allowed for stability.} As the different parameters involved in our model play different roles, the way their variability propagates to $\RelContactsCrit$ differs, even when their variability profiles look similar. 
        \textbf{A:}  Impact of  single-parameter variation on the critical (maximal, allowed) level of contagious contacts $\RelContactsCrit$. To evaluate the monotony (direction) of their impact on $\RelContactsCrit$, we scan their entire definition range, ignoring the practical feasibility of achieving such values. The solid red line shows the default critical level of contacts allowed for stability.
        \textbf{B:} Univariate uncertainties of TTI parameters modeled by beta distributions centered on their default value, and the resulting distribution of critical fraction of contagious contacts $\RelContactsCrit$ (right column). Results are shown assuming testing only (light colors) or testing and tracing (dark colors). Solid red lines represent the default value of $\RelContactsCrit$ in the regime of available TTI capacity.
        \textbf{C:} Distribution of the critical fraction of contacts arising from multivariate uncertainty propagation given by the joint of the distributions shown in (A) for testing only (light colors) or testing and tracing (dark colors). Solid red lines represent the default value of $\RelContactsCrit$ in the regime of available TTI capacity. Results show averages of \num{100000} realizations. 
        }
    \label{fig:UncertaintyPropagation}
\end{figure}

To explore the uni-variate impact different signature parameters have on $\RelContactsCrit$, we studied the zeros of \eqref{eq:nonlinearEig} as a function of $\RelContacts$ using the \texttt{@fzero} MATLAB function (\figref{fig:UncertaintyPropagation}A). Using the same routines and a random sampling procedure, we propagate uncertainties in the values of these parameters, uni-variate  (\figref{fig:UncertaintyPropagation}B), and multivariate (\figref{fig:UncertaintyPropagation}C).

\begin{table}[!h]\caption{Parameter uncertainty propagation. $\alpha$ and $\beta$ are the shape parameters of the beta distribution.}
\label{tab:Parameters_uncertainty}
\centering
\begin{tabular}{l p{5cm} lll l l lll}\toprule
Parameter       & Meaning                       & Median & 95$\%$ CI    &$ \alpha$ & $\beta$ & Dist.&Units\\\midrule
$\xi$           & Asymptomatic ratio            & 0.32   & 0.23--0.42   &     27.5     & 27.8 &beta & \SI{}{-}      \\
$\lambda_s$     & Symptom-driven test rate      & 0.25   & 0.20--0.31   &     56     & 168 &beta & \SI{}{days^{-1}}      \\
$\nu$           & Registered contacts (quarantined)   & 0.07   & 0.03--0.13   &    8.25 & 101.8 & beta & \SI{}{-}       \\
$\eta$          & Tracing efficiency            & 0.66   & 0.59--0.73   &   117.9 & 60.7 & beta & \SI{}{-}       \\
$\epsilon$      & Lost contacts (quarantined)      & 0.05   & 0.01--0.11   &   3.8 & 71.25 & beta & \SI{}{-}       \\\midrule
$\RelContactsCrit\Big|_{\text{TTI}}$      & Maximum --critical-- level of contacts allowed for stability (with TTI)& \SI{61}{\percent}    & 47--\SI{76}{\percent} &   \SI{}{-} & \SI{}{-} & \SI{}{-} & \SI{}{-}       \\
$\RelContactsCrit\Big|_{\text{no TTI}}$   & Maximum --critical-- level of contacts allowed for stability (without TTI)          & \SI{42}{\percent}      & 38--\SI{47}{\percent}       &   \SI{}{-} & \SI{}{-} & \SI{}{-} & \SI{}{-}       \\\bottomrule
\end{tabular}
\end{table}

\begin{table}[h!]\caption{Linearly-derived correspondence between contact reduction and the observed reproduction number}
\label{tab:kt_to_robs}
\centering
\begin{tabular}{lll}\toprule
$\RelContacts$ & $\Rtobs$ (with TTI) & $\Rtobs$ (without TTI)\\\midrule
0.8  & 1.08 & 1.26  \\
0.6  & 0.99 & 1.13  \\
0.4  & 0.91 & 0.98 \\
0.25 & 0.86 & 0.88\\\bottomrule
\end{tabular}
\end{table}

\newpage
\subsection{On the contact reduction required for achieving early population immunity.}\label{sec:HerdImmunity}

In Supplementary Section~\ref{sec:linearStab}, we derived a methodology for obtaining the minimal, critical contact reduction $\RelContactsCrit$  for which the linear system is asymptotically stable. Such values, however, assume a fully susceptible population, as we ignore the scaling factor $\SM$. 

The population immunity threshold $\varrho$ represents the fraction of the population that needs to be immunized for controlling the spread of an infectious disease. It can be expressed in terms of the effective reproduction number $R_t$:

\begin{equation}\label{eq:herd_immunity}
    \varrho = 1-\frac{1}{R_t}.
\end{equation}

In the context of our model, $R_t$ can be expressed in terms of the fraction of contagious contacts $\RelContacts$ and the basic reproduction number $R_0$; $R_t = \RelContacts R_0$. However, in further stages of an ongoing outbreak, the fraction of people no longer susceptible would affect the spread. Thus we include also the scaling factor $\SM$:

\begin{equation}\label{eq:Rt}
    R_t = \RelContacts\SM R_0.
\end{equation}

Combining both equation~\eqref{eq:herd_immunity} and~\eqref{eq:Rt}, we can express $\varrho$ as

\begin{equation}\label{eq:herd_immunityRt}
    \varrho = 1-\frac{1}{\RelContacts\left(1-f\right) R_0},
\end{equation}

assuming a quasi-stationary dynamics for $S$, and defining $f= 1-\SM$. Suppose we study the case in which no major behavioral changes take place. Thus the population immunity threshold $\varrho$ would remain the same. On the other hand, because of the sole fact of having a progressively increasing immunization among the population (because of vaccination or post-infection immunity), the maximal allowed level of contacts $\RelContactsCrit$ will increase. Assuming critical conditions, we use $\RelContacts=\RelContactsCrit$ in equation~\eqref{eq:herd_immunityRt}:

\begin{equation}\label{eq:herd_immunity_crit}
    \varrho = 1-\frac{1}{\RelContactsCrit\left(1-f\right) R_0},
\end{equation}

As we assumed that no behavioral change is taking place, we obtain the population immunity threshold by only evaluating equation~\eqref{eq:herd_immunity_crit} at $f=0$.

\begin{equation}\label{eq:herd_immunity_crit0}
    \varrho = 1-\frac{1}{k^{\rm crit,0} R_0},
\end{equation}

where $k^{\rm crit,0}$ represents the critical level of contacts allowed in a fully susceptible population, and can be calculated directly from the linear stability analysis described in S1.2. Subtracting \eqref{eq:herd_immunity_crit} and \eqref{eq:herd_immunity_crit0} we obtain an expression for $\RelContactsCrit(f)$

\begin{equation}\label{supeq:rcrit_f}
    \RelContactsCrit(f) = \frac{k^{\rm crit,0}}{1-f}.
\end{equation}

Finally, this expression only depends on the remaining susceptible population $(1-f)$, and on the critical level of contacts when $f=0$. Note that this equation can return values of $\RelContactsCrit(f)$ larger than one, should $f$ be \textit{close enough} to one. Following our interpretation of $\RelContactsCrit(f)$, that would mean reaching the population immunity level, as individuals would be allowed to have even higher levels of contacts than those they had before COVID-19.

\subsection{Calculating the increase in the level of contacts allowed with increased immunity \label{vacination}}

In the previous section, we demonstrated that with increased immunity (acquired by vaccination or post-infection), the maximum --critical-- level of contacts $\RelContactsCrit$ allowed for stability would increase. We also discussed that, as a first-order approach, the level at which the system is stabilized (namely, $\Nequil$) is determined by the influx $\Phi_t$  and the distance between the current level of contacts $k_t$ and $\RelContactsCrit$ according to the formula

\begin{equation}\label{supeq:equilibrium}
    \Nequil = \frac{\Phi_t}{\RelContactsCrit-\RelContacts}.
\end{equation}

We can re-write equation~\eqref{supeq:equilibrium} as:

\begin{equation}\label{supeq:rel_contacts_allowed}
    \RelContacts =\frac{k^{\rm crit,0}}{1-f} -  \frac{\Phi_t}{\Nequil},
\end{equation}
where $\RelContacts$ represents the \textit{allowed} level of contagious contacts so that the distance to the singularity (reached if $\RelContacts = \RelContactsCrit$) is kept constant. According to equation~\eqref{supeq:rel_contacts_allowed}, increasing levels of immunity will lead to higher critical level of contacts, thus allowing individuals to steadily increase the level of contagious contacts $\RelContacts$ while still keeping the same equilibrium value for case numbers. 

In order to connect this quantity to the predominant circulating variants, we recall the definition of the hidden reproduction number $R_t^H$ (which is slightly different from the effective reproduction number $R_t$ defined in equation~\eqref{eq:Rt}). As described in the main text and in \cite{contreras2021challenges}, $R_t^H$ accounts for the number of offspring infections generated by individuals unaware of being infectious, in a fully susceptible population. In that way, we can express it in terms of $\RelContacts$; $R^{H}_t = \RelContacts R_0$.  Thus, we can rewrite equation~\eqref{supeq:rel_contacts_allowed} in terms of the allowed hidden reproduction number $R_t^H$, just by multiplying it by $R_0$.

\begin{equation}
    R_t^{H} = \frac{k_t^{\rm crit,0}R_0}{1-f}-\frac{\Phi_t R_0}{\Nequil},
\end{equation}

\begin{figure}[!h]
    \centering
    \includegraphics[width=15cm]{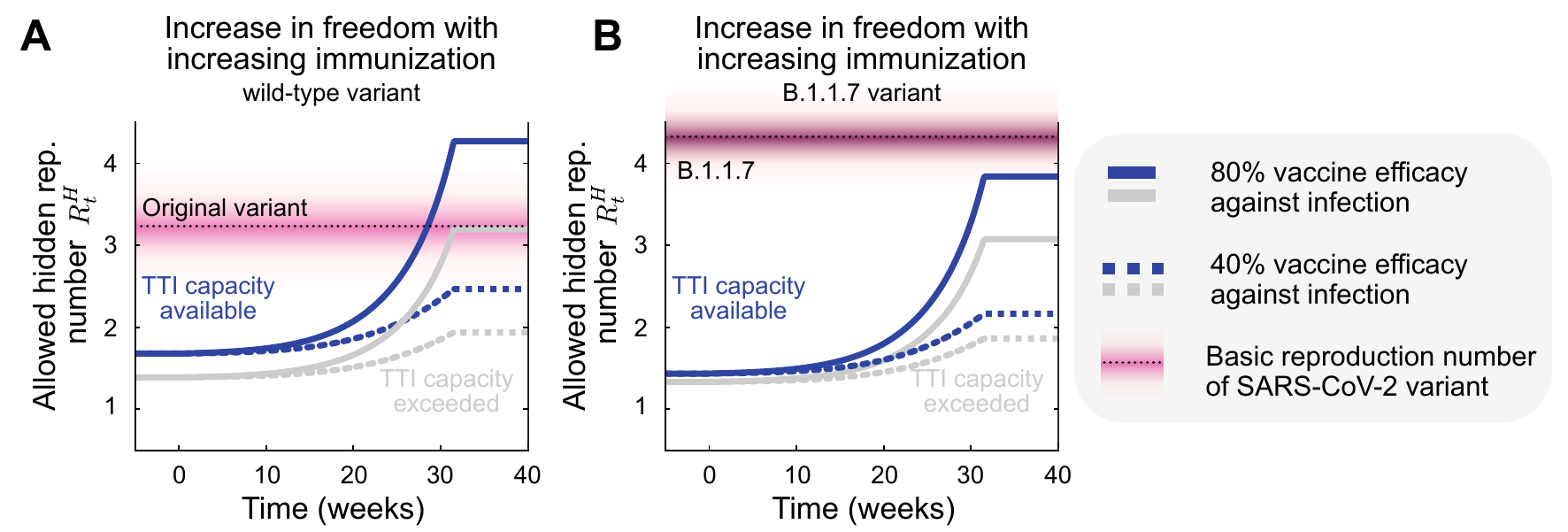}
    \caption{
    \textbf{As growing immunity increases the maximal allowed number of contacts, freedom will proportionally increase.} Scenarios considering the original, wild-type variant of SARS-CoV-2 ($R_0=3.3$) show that vaccination together with a strategy aiming at low case numbers can make the allowed hidden reproduction number $R_t^H$ cross the variant's $R_0$, i.e., the population immunity threshold (\textbf{A}). However, considering the higher reproduction number of the B.1.1.7, restrictions are not allowed to be lifted fully to keep case numbers under control, even if within TTI capacity (\textbf{B}). The vaccination progress here was adapted from Figure~\ref{fig:vaccination} from the main text.
    }
    \label{extfig:RtH_over_time}
\end{figure}

We note a slight coupling between the hidden reproduction number $R_t^{H}$ and the dominant variant of SARS-CoV-2 (through the base reproduction number), so scenarios need to be analyzed separately. Using the vaccination progress described in Fig.~5 in the main text, our results are summarized in~\figref{extfig:RtH_over_time}, where we assume that the TTI and hospital strategies (resp. blue and gray curves) stabilize at 10 and 250 cases per million people, respectively. The influx term that reduces the allowed hidden reproduction number $R_t^H$ highlights the need of combining NPIs aiming to lower case numbers to the TTI regime (where $k_t^{\rm crit,0}$ is higher), especially when large fractions of the population remain susceptible (cf.~\figref{extfig:RtH_over_time}A). However, in the presence of highly contagious variants, NPIs must remain in place, as even after vaccine rollout the allowed hidden reproduction number could not surpass the base reproduction number of the variant. Even though reducing the influx could also help, its effect is little compared with the effect of immunization (cf.~\figref{extfig:RtH_over_time}B).

\subsection{Inferring the reproduction number of COVID-19 from Jun to Oct. 2020}

We use the Bayesian inference framework on an SIR model presented in our previous work~\cite{dehning2020inferring} to infer the daily growth rate $\lambda^*(t)$. An SIR compartmental model with weekly change points is used whose main epidemiological parameters are inferred using the PyMC3 package\cite{salvatier_probabilistic_2016}. A weekly modulation was applied to take the weekly reporting structure and weekend delays into account. After inference, a  rolling average was performed on the daily case numbers for comparability and clarity. The observed reproduction number $\Rtobs$ can be expressed depending on the effective daily growth rate:
\begin{equation}
    \Rtobs = (\lambda^*(t)+1)^4
\end{equation}

\begin{figure}[!h]
    \centering
    \includegraphics[width=10cm]{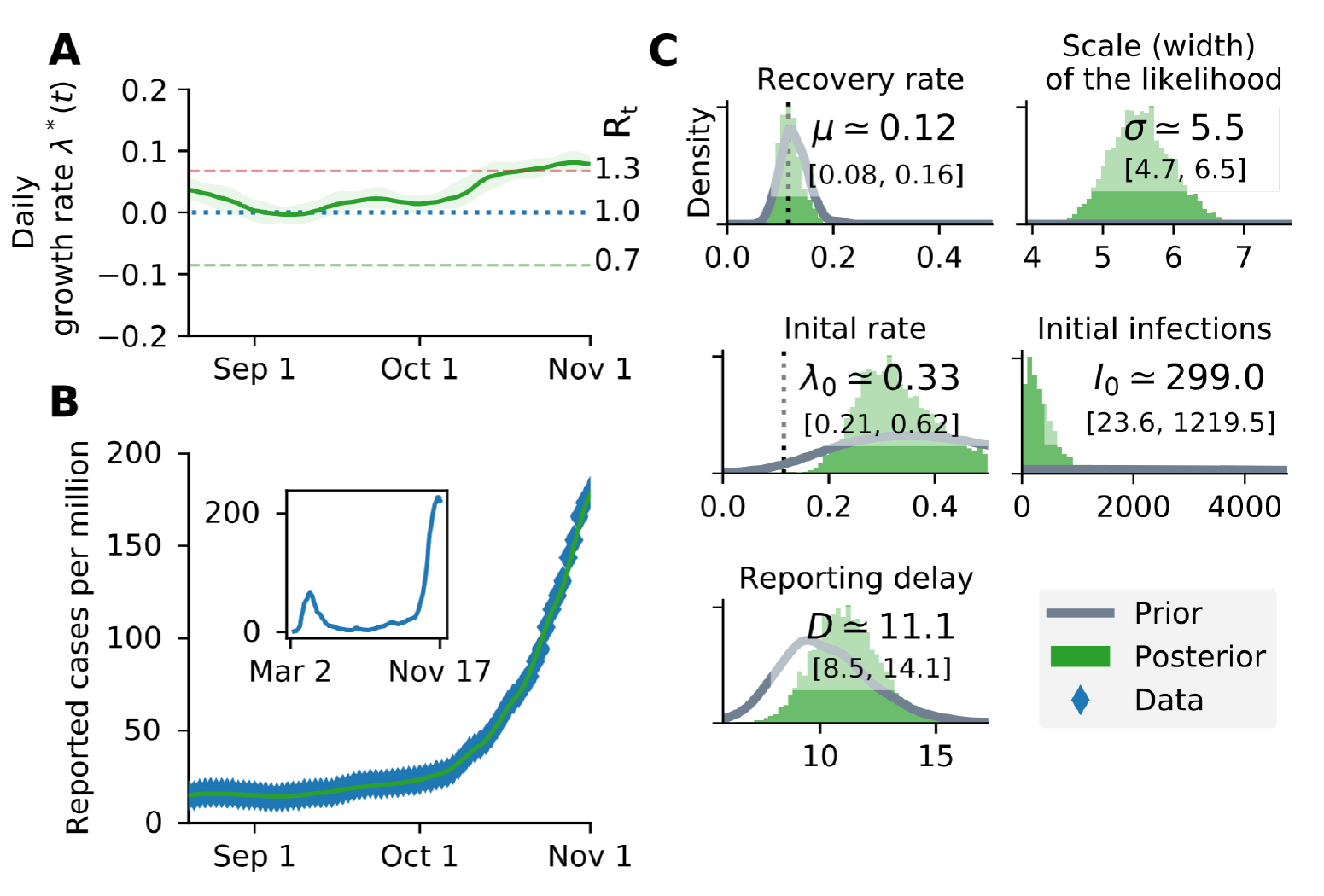}
    \caption{
    \textbf{A substantial increase in Germany's effective growth rate occurred during October, suggesting that regional TTI capacity limits were exceeded.}
    (\textbf{A}) Before October, the reproduction number was slightly above one (corresponding to a daily growth rate $\lambda^\ast$ slightly above zero).
    (\textbf{B}) Observed case numbers were stabilized below 20 daily new cases per million (but still slowly growing) until a transition into the unstable regime took place over $\sim4$ weeks in October.
    The time range is adjusted to focus on this tipping point.
    The inset shows case numbers for the full available time range. (\textbf{C}) Main central epidemiological parameters with the prior and posterior distribution.}
    \label{extfig:Germany}
\end{figure}

\begin{figure}[!h]
    \centering
    \includegraphics[width=15cm]{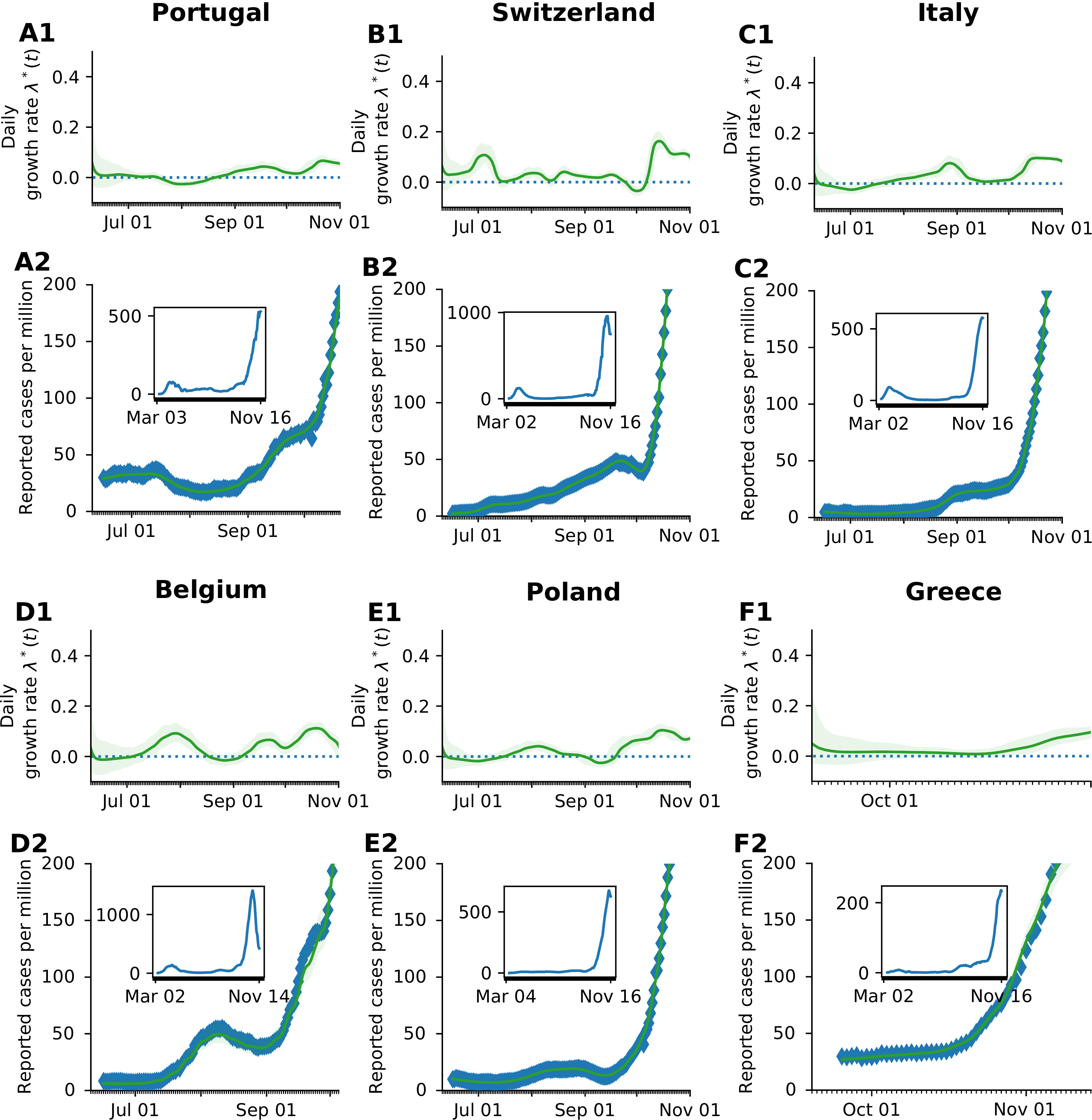}
    \caption{%
        \textbf{Comparison of the reproduction number and reported cases as the second wave emerges in different European countries.} For each country, parameters of an SIR model, were fitted to the reported data of the Our World in Data repository~\cite{owidcoronavirus}, following the procedure presented in~\cite{dehning2020inferring}. 
        (\textbf{Panels X1}) The time-dependent effective growth rate stays between $-0.1$ and $0.1$ and rises before the tipping point. This corresponds to an effective reproduction number between $0.7$ and $1.3$, which matches our preliminary assumptions. The time range is adjusted to focus on the tipping point.
        (\textbf{Panels X2}) After a (meta-)stable regime in summer, all selected countries show a rise in case numbers and a tipping point at around 50 new cases per day per million. The spread self-accelerates, and the cases increase significantly.
        (\textbf{Insets}) Case numbers for the full available time range.
        }
    \label{extfig:europe_bayesian}
\end{figure}

This short analysis was performed for Germany (\figref{extfig:Germany}) and other European countries (\figref{extfig:europe_bayesian}), which showed the same metastable behavior, while their case numbers were below a threshold of around 50 cases per day (per million). A transition into the unstable regime can be seen once case numbers surpass this threshold.

\subsection{On the incorporation of random testing in the TTI scheme}
 
Testing can also be done randomly or randomly combined with a contact tracing strategy. Even though the number of tests required for such purposes would be enormous, the development of fast, cheap, and reliable tests offers an exciting alternative to consider. 

In this supplementary note, we derive the equations presented in the methods but also including random testing. As therein described, random testing is assumed to occur at a constant rate $\lambda_r$, which, for default parameters, reflect the number of tests per day per million people. In that way, it seems reasonable to consider unfeasible testing rates surpassing $\lambda_r^{\rm max} = 0.1$, as it would mean that a \SI{10}{\%} of the population is tested every day. 

\subsubsection{Number of cases observable through testing \texorpdfstring{$\ftestH$}{Ntest}}

When random testing is included in the scheme, the solution of equation~\eqref{eq:dYds} for the symptomatic and asymptomatic infections -hidden- would be given by equations~\eqref{supeq:chisr} and~\eqref{supeq:chir}.
\begin{align}
\chisr&=
\begin{cases}
     &\dis\latRate\tau\exp\left(-\dis\latRate\tau\right), \quad \text{if }  \latRate\approx\lambda_s+\lambda_r \\                       &\dis\frac{\latRate}{(\lambda_s+\lambda_r)-\latRate}\left(\exp\left(-\dis\latRate\tau\right)-\exp\left(-\tau\left(\lambda_s+\lambda_r\right)\right)\right),                     \text{else}. \label{supeq:chisr}\\
\end{cases}
\\
\chir&=
\begin{cases}
    &\dis\latRate\tau\exp\left(-\dis\latRate\tau\right), \quad \text{if }  \latRate\approx\lambda_r \\                                 &\dis\frac{\latRate}{\lambda_r-\latRate}\left(\exp\left(-\dis\latRate\tau\right)-\exp\left(-\tau\lambda_r\right)\right), \text{else}. \label{supeq:chir}\\
\end{cases}
\end{align}

If both symptom-based and random testing take place simultaneously, the number of discovered infections is given by 

\begin{equation}
     \ftestH  = \lambda_r\Ht + \lambda_s \Hts
\end{equation}

Further, assuming that after reaching $\Nmax$, the testing rates at the overhead pool-sizes would decrease to $\lambda_s'$ and $\lambda_r'$, respectively, for symptom-driven and random testing. We further assume that testing resources would be exclusively allocated to sustain the symptom-driven testing in our default scenario. The overall testing term $\ftestH$ would be given by:

\begin{equation}
\begin{split}
    \ftestH  =  \lambda_r\min\left(\Ht,\Htmax\right) & + \lambda_r'\max\left(0,\Ht\!-\!\Htmax\right) \\ &+ \lambda_s\min\left(\Hts,\Htsmax\right) + \lambda_s'\max\left(0,\Hts\!-\!\Htsmax\right),
\end{split}
\end{equation}

where $\Htsmax,\Htmax$ represent the pool sizes of the symptomatic hidden and total hidden pools, respectively, at the TTI limit, i.e. $\lambda_r\Htmax + \lambda_s\Htsmax \overset{!}{=} \Nmax$, reached at time $t=t^*$. Defining $\varphi:=\dis\frac{\Hts}{\Ht}\Big|_{t=t^{*}}$, we can express such magnitudes in term of the maximum capacity $\Nmax$:

\begin{eqnarray}
    \Htsmax & =  \dis\frac{\varphi\Nmax}{\varphi\lambda_s+\lambda_r}\\
    \Htmax & =  \dis\frac{\Nmax}{\varphi\lambda_s+\lambda_r}.
\end{eqnarray}

The explicit value of $\varphi$ can be obtained numerically in the integration routine or estimated through the use of the equilibrium values of the differential equations, $\varphi = \frac{\Hts_\infty}{\Ht_\infty}$ (as implemented in our code). The expression for the symptomatic hidden pool $\Hts$ in the presence of random testing is slightly different; 

\begin{equation}
\begin{split}
    \ftestHs  =  \lambda_r\min\left(\Hts,\Htsmax\right) & + \lambda_r'\max\left(0,\Hts\!-\!\Htsmax\right) \\ &+ \lambda_s\min\left(\Hts,\Htsmax\right) + \lambda_s'\max\left(0,\Hts\!-\!\Htsmax\right).
\end{split}
\end{equation}

\subsubsection{Random testing: number of cases observable through contact-tracing \texorpdfstring{$\ftrace$}{Ntraced}}

Because of TTI, infectious individuals move (or are likely to move) from the hidden to the quarantined infectious pool before recovering. Therefore, they spend a comparatively shorter amount of time there and, on average, would not generate the expected amount of offspring infections as some would be prevented. 
In the absence of TTI, the average time individuals spend in the infectious compartment is $\frac{1}{\gamma}$. 
In the presence of selective TTI, symptomatic individuals would have a greater chance to be tested (and thereby removed) than the asymptomatic ones. As we explicitly consider compartments for symptomatic and asymptomatic infections, each pool's residence times would be different. Noting that symptomatic individuals can be tested and therefore removed by any of the testing criteria, their residence time would be approximate $\frac{1}{\gamma + \lambda_s + \lambda_r}$. In contrast, the average residence time of asymptomatic individuals would be  $\frac{1}{\gamma+\lambda_r}$. Therefore, the fractions of time that symptomatic and asymptomatic individuals stay unnoticed are respectively

\begin{equation}
    t_s = \frac{\gamma}{\gamma + \lambda_s + \lambda_r},
\end{equation}

and

\begin{equation}
    t_r = \frac{\gamma}{\gamma + \lambda_r}.
\end{equation}

If the daily new cases observed through testing, delayed at the moment of processing, $\Ntestlag$, are within the tracing capacity of the health authorities, i.e. $\Ntestlag \leq \Nmax$, then $\ftrace$ is defined as

\begin{equation}
    \ftrace =  \eta R_{t-\tau}\left(\Htlag t_r \lambda_r+\Hslag \left(t_s\lambda_s+\left(t_s-t_r\right)\lambda_r\right)\right),
\end{equation}

where $R_{t-\tau}$ represents the effective reproduction number, as defined in equation~\eqref{eq:Rt}. Otherwise, using the expressions for $\Ht$ and $\Hts$ when the TTI capacity is reached derived in the previous section, we can obtain an effective rate

\begin{equation}
    \lambda_{\rm eq} =  \frac{ \lambda_r\left(\left(1-\varphi\right) t_r + \varphi t_s \right)+ \varphi \lambda_s t_s}{\lambda_r+\varphi\lambda_s}.
\end{equation}

Therefore, the average amount of positive cases identified by contact tracing in the TTI limit is given by

\begin{equation}
    \ftrace =  \eta R_{t-\tau} \Nmax \lambda_{\rm eq}.
\end{equation}

To sum up the last equations:
\begin{align}
\ftrace=
\begin{cases}
    \eta R_{t-\tau} (\Htlag t_r \lambda_r+\Hslag \left(t_s\lambda_s+\left(t_s-t_r\right)\lambda_r\right)) \quad &\text{if } \Ntestlag \leq \Nmax\\
    \eta R_{t-\tau} \Nmax \dis \lambda_{\rm eq} & \text{else}
\end{cases}
\end{align}

\subsection{More analytical insights into the TTI-based metastable regime}

One of the crucial aspects of this paper is the description and profit from a metastable regime at low case numbers. Because of the different compartments at interplay in the dynamics, some metastability aspects are easier to present using a simplified model. The critical element behind the case-number dependent metastability is the limited capacity to perform contact tracing of the newly identified infections. We refer to the TTI limit being reached at a prevalence given by $I=I^{\rm max}$. However, the fact that this metastable equilibrium shows non-zero case numbers arises from a small but non-vanishing influx of externally acquired infections to the system, which we call $\Phit$. 

Aiming to illustrate the dynamics in a minimalist way, we proceed as follows. We modify a plain SIR model to include hidden and quarantined infections but simplify the TTI scheme to only depend on two parameters: the TTI capacity referred to the hidden incidence $I_H^{\rm max}$ and the testing rate $\lambda$ as long as the capacity is not exceeded. If the capacity is exceeded the testing rate drops to zero. Infections in the quarantined and hidden pools spread at rates $\beta_1$ and $\beta_2$ respectively, with $\beta_1<\beta_2$, and the recovery rate is given by $\gamma$:

\begin{align}
\frac{dS}{dt} & = - \beta_1 I^Q \frac{S}{M}  - \beta_2 I^H \frac{S}{M} - \Phit\\
\frac{dI^Q}{dt} & = \beta_1 I^Q \frac{S}{M} - \gamma I^Q + \min(\lambda I^H, \lambda I^H_{\rm max}) \\
\frac{dI^H}{dt} & = \beta_2 I^H \frac{S}{M} - \gamma I^H - \min(\lambda I^H, \lambda I^H_{\rm max}) + \Phit. 
\end{align}

We observe that the metastable regime's key ingredients relate to TTI: The system reacts less intensively to new infections when case numbers are below the TTI capacity, as infection chains are diligently found and removed. On the other hand, when the TTI system is overwhelmed, infections reproduce at the natural, hidden rate. Assuming quasi-stationary dynamics for the susceptible pool ($S$), we can approximate:

\begin{equation}
   \frac{d }{d I^H} \left(\frac{dI^H}{dt}\right)  =\begin{cases}
     &\dis\beta_2 \frac{S}{M} - \gamma - \lambda, \quad \text{if }  I^H < I^H_{\rm max}, \\ & \\                      &\dis\beta_2 \frac{S}{M} - \gamma, \quad                     \text{otherwise}.\\
\end{cases}
\end{equation}

If $I^H$ exceeds $I^H_{\rm max}$, there is the possibility for $I^H$ to grow and lead to a more significant number of cases for a broader range of parameters. On the other hand, when studying the linear system's equilibrium, we identify the contributions to the observed number of cases $\Nequil = \beta_1 I^{Q}_{\infty} + \lambda I^{H}_{\infty}$. Assuming stationary conditions for the $\SM\approx 1$ limit, we obtain:

\begin{equation}
    \Nequil = \frac{\Phit}{\lambda+\gamma-\beta_2}\frac{\gamma\lambda}{\gamma-\beta_1}.
\end{equation}

Using these simple models, we can better understand the parameters and mechanisms that render the system unstable or the equilibrium unfeasible. This better understanding of factors contributing to stabilizing --and respectively destabilizing-- the system, helps to guide policies. The transitions and equilibrium levels have more complex analytical expressions in our model but follow the same spirit: to draw the line between the stable and unstable spread.

\subsection{Exploring the effect of more compartments for the exposed individuals}

\begin{figure}[!h]
    \centering
    \includegraphics[width=15cm]{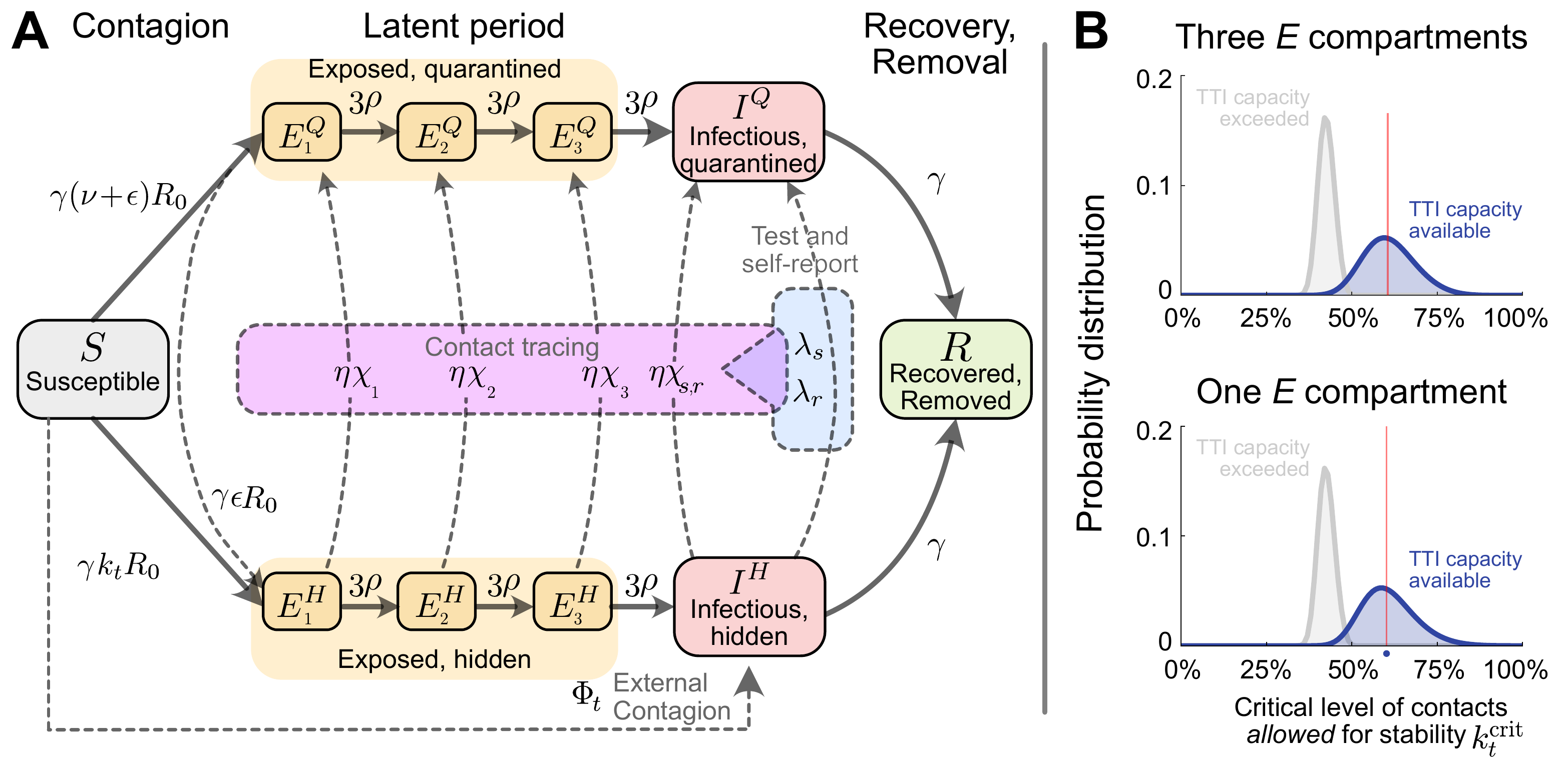}
    \caption{%
        \textbf{Flowchart of the complete model, including three compartments for the latent period.} \textbf{A:} The solid blocks in the diagram represent different SEIR compartments for both hidden and quarantined individuals. Hidden cases are further divided into symptomatic and asymptomatic carriers (not shown).
        Solid lines represent the natural progression of the infection (contagion, latent period, and recovery). On the other hand, dashed lines account for imperfect quarantine and limited compliance, external factors, and test-trace-and-isolate policies. \textbf{B:} Comparison between the distribution of $\RelContactsCrit$ for the single-compartment model and the three compartments model.}
    \label{extfig:three_compartments}
\end{figure}

Compartmental models coupled with differential equations implicitly assume first-order kinetics, which translates to exponential emptying or filling the compartments involved in the dynamics. Even though the expected residence time can be set for each compartment, the exponential shape of the residence time distribution differs from the true disease dynamics.
This difference can be corrected by including extra compartments, so instead of having an exponential shape, the residence time distribution eventually converges to a delta distribution around the desired value. Aiming to evaluate how this effect might change our estimates for $\RelContactsCrit$, we included three compartments for the exposed fraction of the population (and for hidden and quarantined infections) presented in~\figref{extfig:three_compartments}. As the latent period would be spent in these compartments, the residence time in each one is one-third of the original, so that the new transition rate would be $\latRate' = 3\latRate$.

Following the same formalism presented in the previous sections, we estimate the fraction of individuals infected at time $t$ that would remain in the different compartments by the time of contact tracing. 

\begin{align}
    \chi_1 & = \exp\left(-3\latRate\tau\right),\\
    \chi_2 & = \left(3\latRate\right)\tau\exp\left(-3\latRate\tau\right),\\
    \chi_3 & = \left(3\latRate\right)^2\frac{\tau^2}{2}\exp\left(-3\latRate\tau\right).
\end{align}

Consistently, we can demonstrate that the fraction of individuals staying in the infectious compartment in the presence of symptom-based testing is given by

\begin{equation}
\chisr=
\begin{cases}
     &\dis\left(3\latRate\right)^3\frac{\tau^3}{3!}\exp\left(-3\latRate\tau\right), \quad \text{if }  3\latRate\approx\lambda_s+\lambda_r \\                       &\dis\frac{\left(3\latRate\right)^3\exp\left(-\dis3\latRate\tau\right)}{2(\lambda_s+\lambda_r-3\latRate)}\left(\tau^2-2\frac{\tau}{\lambda_s+\lambda_r-3\latRate}+\frac{2}{\left(\lambda_s+\lambda_r-3\latRate\right)^2}\left(1-\exp\left(-\tau\left(\lambda_s+\lambda_r-3\rho\right)\right)\right)\right),                     \text{else}.\\
\end{cases}
\end{equation}

To analyze the stability of the extended system with three exposed compartments, we proceeded as described in Section~\ref{sec:linearStab}. The linear system with delay representing the dynamics is $x'(t) = Ax(t)+Bx(t-\tau)$, with $x(t)=\left[\ET_1(t);\,\ET_2(t);\,\ET_3(t);\,\EH_1(t);\,\EH_2(t);\,\EH_3(t);\,\Tt(t);\,\Ht(t);\,\Hts(t)\right]$, where matrices $A$ and $B$ are given by:

\begin{eqnarray}
    A & = & 
    \begin{pmatrix} 
    -3\latRate  & 0 & 0 & 0 & 0 & 0                     & \nu\gamma R_0         & 0                         & 0 \\
     3\latRate  & -3\latRate & 0 & 0 & 0 & 0            & 0    & 0   & 0 \\
    0           &  3\latRate  & -3\latRate & 0 & 0 & 0 & 0               & 0    & 0 \\
    0 & 0 & 0 & -3\latRate & 0 & 0            & \epsilon\gamma R_0    & \gamma\RtH    & 0 \\
    0 & 0 & 0 & 3\latRate & -3\latRate & 0            & 0   & 0    & 0 \\
    0 & 0 & 0 & 0 & 3\latRate & -3\latRate            & 0    & 0    & 0 \\
    0 & 0 & 3\latRate &0 &0 &0                                       & -\gamma               & \lambda_r                 & \lambda_s \\
    0 & 0 & 0 & 0 & 0 & 3\latRate                              & 0                     & -\gamma-\lambda_r         & -\lambda_s\\
    0 & 0 & 0 & 0 & 0 & 3\left(\!1-\!\xi\!\right)\latRate      & 0                     &        0                  & -\gamma -\lambda_r -\lambda_s \\
    \end{pmatrix}\\
    B & = & 
    \begin{pmatrix} 
    0   & 0 & 0 & 0   & 0 & 0 & \lambda_r^{\rm eff} \chi_1   & \lambda_s^{\rm eff} \chi_1    \\
    0   & 0 & 0 & 0   & 0 & 0 & \lambda_r^{\rm eff} \chi_2   & \lambda_s^{\rm eff} \chi_2  \\
    0   & 0 & 0 & 0   & 0 & 0 & \lambda_r^{\rm eff} \chi_3   & \lambda_s^{\rm eff} \chi_3    \\
    0   & 0 & 0 & 0   & 0 & 0 & -\lambda_r^{\rm eff} \chi_1  & -\lambda_s^{\rm eff} \chi_1  \\
    0   & 0 & 0 & 0   & 0 & 0 & -\lambda_r^{\rm eff} \chi_2  & -\lambda_s^{\rm eff} \chi_2    \\
    0   & 0 & 0 & 0   & 0 & 0 & -\lambda_r^{\rm eff} \chi_3  & -\lambda_s^{\rm eff} \chi_3  \\
    0   & 0 & 0 & 0   & 0 & 0 &\lambda_r^{\rm eff}  \left(\xi\chir + \left(\!1-\!\xi\!\right)\chisr\right)  & \lambda_s^{\rm eff}  \left(\xi\chir + \left(\!1-\!\xi\!\right)\chisr\right)               \\
    0   & 0 & 0 & 0   & 0 & 0 & -\lambda_r^{\rm eff} \left(\xi\chir + \left(\!1-\!\xi\!\right)\chisr\right) & -\lambda_s^{\rm eff} \left(\xi\chir + \left(\!1-\!\xi\!\right)\chisr\right)\\
    0   & 0 & 0 & 0   & 0 & 0 & -\lambda_r^{\rm eff} \left(\!1-\!\xi\!\right)\chisr                      &   -\lambda_s^{\rm eff} \left(\!1-\!\xi\!\right)\chisr\\
    \end{pmatrix}\eta\RelContacts R_0,
\end{eqnarray}

We obtained new estimates for $\RelContactsCrit$ in the cases of available TTI capacity and overwhelmed TTI capacity. Comparing these values to those obtained in our original model (with a single exposed compartment), we find that they do not deviate significantly. Graphically, we see that the distributions share the same properties, being slightly less skewed in the three-compartment case (see~\figref{extfig:three_compartments}B). In numbers, the relative error we induce in calculating the critical level of contacts allowed for stability by using a single exposed compartment instead of three is reported in table \ref{tab:rel_error_three_compartments}.

\begin{table}[h!]\caption{Linearly-derived correspondence between contact reduction and the observed reproduction number (7.5e4 realizations).}
\label{tab:rel_error_three_compartments}
\centering
\begin{tabular}{lllll}\toprule
            & & \multicolumn{3}{c}{Percentile}\\\midrule
\multicolumn{2}{c}{Variable}    & \SI{2.5}{\%} & \SI{50}{\%} (median) & \SI{97.5}{\%}\\\midrule
\multirow{2}{*}{$\RelContactsCrit\Big|_{\text{no TTI}}$} & three compartments  & \SI{37.80}{\%}  & \SI{42.58}{\%} & \SI{46.68}{\%}  \\
                                                         & one compartment     & \SI{37.71}{\%}  & \SI{42.59}{\%} & \SI{46.71}{\%}  \\\midrule
\multirow{2}{*}{$\RelContactsCrit\Big|_{\text{TTI}}$} & three compartments  & \SI{47.19}{\%}  &\SI{60.60}{\%} & \SI{76.39}{\%}  \\
                                                         & one compartment     & \SI{46.99}{\%}  & \SI{60.48}{\%} & \SI{76.20}{\%}  \\\midrule
\multirow{2}{*}{Relative error} & no TTI  & \SI{0.22}{\%}  & \SI{0.03}{\%}& \SI{0.07}{\%}  \\
                                                         & TTI     & \SI{0.42}{\%} & \SI{0.19}{\%} & \SI{0.24}{\%}  \\
                                                         \bottomrule
\end{tabular}
\end{table}

\end{document}